\documentclass[sigconf]{acmart}

\usepackage{amssymb,amsfonts}
\usepackage{textcomp}
\usepackage{xcolor}
\usepackage{tabularray}%
\usepackage{svg}
\usepackage[normalem]{ulem}%
\usepackage{makecell}
\usepackage{subcaption}
\usepackage{url}
\usepackage{framed}
\usepackage[skip=1pt,labelfont=bf]{caption}
\usepackage{calc}
\usepackage{color}
\usepackage{colortbl}
\usepackage{enumitem}
\usepackage{fancybox}
\usepackage{fontawesome5}
\usepackage{listings}
\usepackage{lscape}
\usepackage{multirow}
\usepackage{pifont}
\usepackage{rotating}
\usepackage{siunitx}
\usepackage{soul}
\usepackage{xspace}

\definecolor{codegreen}{rgb}{0,0.6,0}
\definecolor{codered}{rgb}{1,0,0}
\definecolor{codegray}{rgb}{0.5,0.5,0.5}
\definecolor{codepurple}{rgb}{0.58,0,0.82}
\definecolor{backcolour}{rgb}{0.95,0.95,0.92}
\definecolor{lightgray}{gray}{0.9}

\newboolean{showcomments}
\setboolean{showcomments}{true}
\ifthenelse{\boolean{showcomments}}
 { \newcommand{\mynote}[2]{
      \fbox{\bfseries\sffamily\scriptsize#1}
        {\small$\blacktriangleright$\textsf{\emph{#2}}$\blacktriangleleft$}}}
        { \newcommand{\mynote}[2]{}}
        
\definecolor{DarkOrange}{rgb}{0.8,0.3,0.0}
\definecolor{DarkCyan}{rgb}{0.0, 0.55, 0.55}
\definecolor{DarkCyel}{rgb}{1.0, 0.49, 0.0}
\definecolor{yellow-green}{rgb}{0.6, 0.8, 0.2}

\newcolumntype{?}{!{\vrule width 1pt}}

\newcommand{\lantern}{\textsc{LANTERN}\xspace}

\definecolor{findgray}{gray}{0.95}

\newenvironment{findingbox}{%
  \def\FrameCommand##1{%
    \setlength{\fboxsep}{6pt}
    \setlength{\fboxrule}{1.5pt}
    \textcolor{black}{\vrule width \fboxrule}
    \colorbox{findgray}{##1}
    \textcolor{black}{\vrule width \fboxrule}
  }%
  \MakeFramed {\advance\hsize-\width \FrameRestore}}%
 {\endMakeFramed}

\newcommand{\findx}[1]{%
  \begin{findingbox}%
    \noindent\em #1
  \end{findingbox}%
}

\copyrightyear{2026}
\acmYear{2026}
\setcopyright{cc}
\setcctype{by}
\acmConference[ICSE '26]{2026 IEEE/ACM 48th International Conference on Software Engineering}{April 12--18, 2026}{Rio de Janeiro, Brazil}
\acmBooktitle{2026 IEEE/ACM 48th International Conference on Software Engineering (ICSE '26), April 12--18, 2026, Rio de Janeiro, Brazil}
\acmPrice{}
\acmDOI{10.1145/3744916.3773227}
\acmISBN{979-8-4007-2025-3/2026/04}

\begin{document}

\title[\resizebox{4.5in}{!}{Unlocking LLM Repair Capabilities Through Cross-Language Translation and Multi-Agent Refinement}]{Unlocking LLM Repair Capabilities Through Cross-Language Translation and Multi-Agent Refinement}


\author{Wenqiang Luo}
\affiliation{%
  \institution{Department of Computer Science, City University of Hong Kong}
  \country{China}
}
\email{wenqialuo4-c@my.cityu.edu.hk}

\author{Jacky Wai Keung}
\affiliation{%
  \institution{Department of Computer Science, City University of Hong Kong}
  \country{China}
}
\email{Jacky.Keung@cityu.edu.hk}

\author{Boyang Yang}
\affiliation{%
  \institution{Jisuan Institute of Technology, Beijing JudaoYouda Network Technology Co. Ltd.}
  \country{China}}
\email{yangboyang@jisuanke.com}

\author{Jacques Klein}
\affiliation{%
  \institution{SnT, University of Luxembourg}
  \country{Luxembourg}}
\email{jacques.klein@uni.lu}

\author{Tegawend\'e F. Bissyand\'e}
\affiliation{%
  \institution{SnT, University of Luxembourg}
  \country{Luxembourg}}
\email{tegawende.bissyande@uni.lu}

\author{Haoye Tian}
\authornote{Corresponding author.}
\affiliation{%
  \institution{Department of Computer Science, Aalto University}
  \country{Finland}
}
\email{haoye.tian@aalto.fi}

\author{Bach Le}
\affiliation{%
  \institution{School of Computing and Information Systems, University of Melbourne}
  \country{Australia}
}
\email{bach.le@unimelb.edu.au}



\renewcommand{\shortauthors}{Luo et al.}

\begin{abstract}
Recent advances in leveraging LLMs for APR have demonstrated impressive capabilities in fixing software defects. 
However, current LLM-based approaches predominantly focus on mainstream programming languages like Java and Python, neglecting less prevalent but emerging languages such as Rust due to expensive training resources, limited datasets, and insufficient community support. 
This narrow focus creates a significant gap in repair capabilities across the programming language spectrum, where the full potential of LLMs for comprehensive multilingual program repair remains largely unexplored. 
To address this limitation, we introduce a novel cross-language program repair approach \lantern that leverages LLMs' differential proficiency across languages through a multi-agent iterative repair paradigm. Our technique strategically translates defective code from languages where LLMs exhibit weaker repair capabilities to languages where they demonstrate stronger performance, without requiring additional training. A key innovation of our approach is an LLM-based decision-making system that dynamically selects optimal target languages based on bug characteristics and continuously incorporates feedback from previous repair attempts.

We evaluate our method on xCodeEval, a comprehensive multilingual benchmark comprising 5,068 bugs across 11 programming languages. Results demonstrate significant enhancement in repair effectiveness, particularly for underrepresented languages, with Rust showing a 22.09\% improvement in $Pass@10$ metrics. 
Our research provides the first empirical evidence that cross-language translation significantly expands the repair capabilities of LLMs and effectively bridges the performance gap between programming languages with different levels of popularity, opening new avenues for truly language-agnostic automated program repair.

\end{abstract}




\maketitle

\section{Introduction}
Software bugs remain an unavoidable challenge in modern software development where automated program repair (APR) \cite{zhang2023survey,gazzola2018automatic, monperrus2018automatic,long2016automatic} has emerged as a promising solution by automatically identifying and generating patches for software defects. Driven by the rapid revolution in Large Language Models (LLMs) with unprecedented capabilities in generating complex code fixes \cite{sobania2023analysis,xia2023automated,fan2023automated,fan2024oracle}, LLM-based APR approaches have demonstrated remarkable success in addressing diverse software defects, from semantic bugs \cite{wei2023copiloting,wang2023rap} and security vulnerabilities \cite{zhou2024out,fu2022vulrepair} to syntax errors \cite{deligiannis2023fixing,ahmed2022synshine} and hardware security issues \cite{ahmad2024hardware,yao2024hdldebugger}.

By benefiting from training or fine-tuning \cite{hossain2024deep,jin2023inferfix,yuan2022circle,luo2024fine} on diverse multilingual codebases, LLM-based repair approaches have achieved promising performance. However, these models exhibit substantial disparities in fixing bugs across languages, with significantly stronger repair capabilities in prevalent languages \cite{madnight,octoverse} such as Java and Python \cite{yuan2022circle,joshi2023repair,xia2022less} while struggling with others like Rust and Kotlin. 

Notably, on the multilingual benchmark xCodeEval \cite{khan2024xcodeeval}, the state-of-the-art LLM initially achieves $Pass@10$ scores of 89.02\% and 89.93\% for Python and PHP, respectively. However, for less common languages such as Rust, the $Pass@10$ score drops dramatically to only 65.58\%, revealing a significant performance gap of over 24\% points. This disparity primarily stems from the imbalanced distribution of training data, where mainstream languages such as Python dominate open source repositories with millions of contributors compared to relatively newer languages like Rust (according to the statistics of GitHub project contributors up to 2024 \cite{octoverse2024}). Furthermore, the study of Zhang et al. \cite{zhang2024systematic} demonstrates that the imbalance in programming languages also persists in current academia, where the majority of LLM-based APR research focuses on languages such as Java, Python, and C, while studies concerning less prevalent ones such as Rust, Go and Ruby constitute only 13\% of the total research as of 2024. The scarcity of mature high-quality datasets, the high cost of training or fine-tuning, relatively less support in the APR community, and diverse language characteristics \cite{zhang2019study} further amplify these challenges, limiting the practicality and scalability of existing approaches. 

Programming problems often have equivalent implementations across different languages that serve the same underlying intentions and functions \cite{roziere2020unsupervised,tehrani2024coderosetta}, which suggests a natural opportunity: if a bug proves difficult to fix in one language, it might be beneficial to draw from the repair expertise available in another language. Intuitively, in other words, an LLM that exhibits superior repair capabilities in one language could be leveraged to assist with resolving issues in another. In addition, as the repair process unfolds new insights and historical experiences \cite{wen2019historical}, these can be fed back into the repair cycle to guide future refinement. Inspired by these intuitions, we propose to leverage cross-language knowledge by translating the buggy code to other languages with iterative refinement using historical feedback.

LLMs trained in one language can sometimes even outperform those trained in the target language itself~\cite{ahmed2022multilingual}, suggesting valuable transfer effects across language boundaries. With rapid progress in code translation \cite{zhu2022multilingual,roziere2020unsupervised}, the study of Pan et al. \cite{pan2024lost} demonstrates that LLMs can translate code better than non-LLM translation approaches for specific languages, demonstrating the potential of LLMs in code translation. Furthermore, benefited from advances in software engineering agents \cite{liu2024large,yang2025swe}, the integration of agent-driven \cite{xia2024automated,zhang2024autocoderover,bouzenia2024repairagent,hidvegi2024cigar} and multi-agent \cite{lee2024unified,zhang2024acfix} approaches can enhance program repair with historical feedback while autonomously navigating complex repair workflows throughout the iterative repair process. Therefore, we introduce the key hypothesis:
\vspace{-1mm}
    
\begin{framed}
\noindent\textbf{``When the LLM fails to repair buggy code in a given programming language,
\uline{translating the code into another language} where the model demonstrates stronger
repair capabilities may facilitate successful bug fixing.''}
\end{framed}
\vspace{-1mm}

\textbf{This work.} In this paper, we present a novel program repair approach by leveraging cross-\textbf{LAN}guage \textbf{T}ranslation and multi-ag\textbf{E}nt \textbf{R}efi\textbf{N}ement (\lantern) that strategically translates bugs to other programming languages with a multi-agent iterative repair paradigm. Instead of directly fixing the faulty code within its original language, \lantern translates the code into a target programming language, which is selected based on the decision-making capabilities of the LLM that reasons about the bug characteristics and historical feedback. Once the bug is translated, repair is attempted in this alternative language, and the successfully repaired code is subsequently translated back to the original language. Repairs that fail to pass the test suites are not discarded, but instead scheduled for subsequent iterations of refinement, enabling a progressive improvement cycle. We conduct comprehensive experiments to evaluate whether and how \lantern\ works. The evaluation results demonstrate that our approach successfully addresses most bugs that resist direct LLM-based repair attempts. Additional ablation studies confirm that improved repair performance stems from cross-language translation rather than simply repeated fix iterations, demonstrating the unique repair opportunities provided by cross-language perspectives.

\textbf{Contributions.} This paper makes the following contributions:
\begin{itemize}[leftmargin=*]
    \item \textbf{[Hypothesis]} We propose a novel hypothesis on enhancing cross-language repair capabilities through language translation.
    \item \textbf{[Technique]} Based on the hypothesis, we propose a program repair approach termed \lantern that analyzes and translates buggy code to other programming languages in a multi-agent paradigm to enhance cross-language program repair through iterative multi-agent refinement. 
    \item \textbf{[Evaluation]} We perform an extensive evaluation on xCode-Eval, which is a state-of-the-art multilingual benchmark that consists of 5068 bugs across 11 programming languages. Evaluation results reveal that our approach can significantly enhance program repair across all languages. Notably, Rust achieves an improvement of 22.09\% on $Pass@10$.
    \item \textbf{[Analysis]} We conduct an ablation study to validate the impact of the translator and analyzer, both of which contribute to performance improvement.
\end{itemize}

\textbf{Availability.} Artifacts of this work including code, dataset, and implementation details are publicly available at \url{https://github.com/stringing/LANTERN}.
 

\section{Background and Related Work}
\subsection{Code Translation}
Code translation refers to the process of transforming source code from one programming language to another, mainly employed for code adaptation scenarios or migrating legacy codebases written in deprecated languages to more modern alternatives \cite{zhu2022multilingual}. 
\begin{figure*}[!t]
  \centering
    \vspace{-3mm}\includegraphics[width=.8\linewidth]{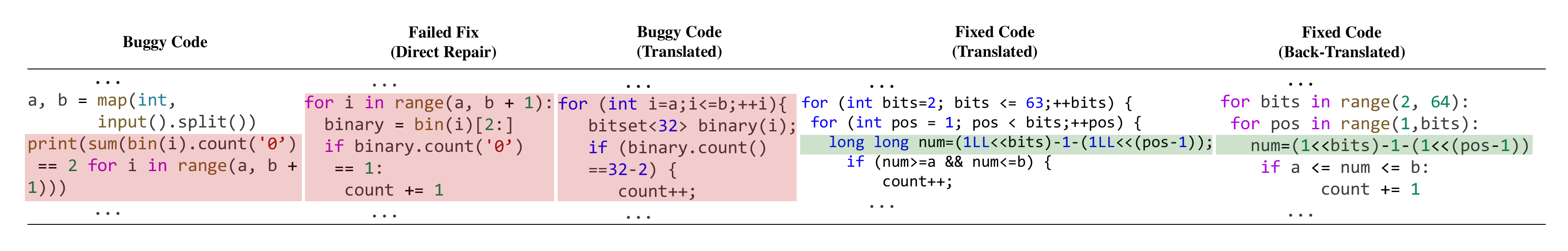}
  \caption{A motivating example of fixing a Python bug by translating it to C++.}
  \label{motivation-1}
    \vspace{-3mm}
\end{figure*}
\begin{figure*}[!t]
\vspace{-3mm}
  \centering\includegraphics[width=.8\linewidth]{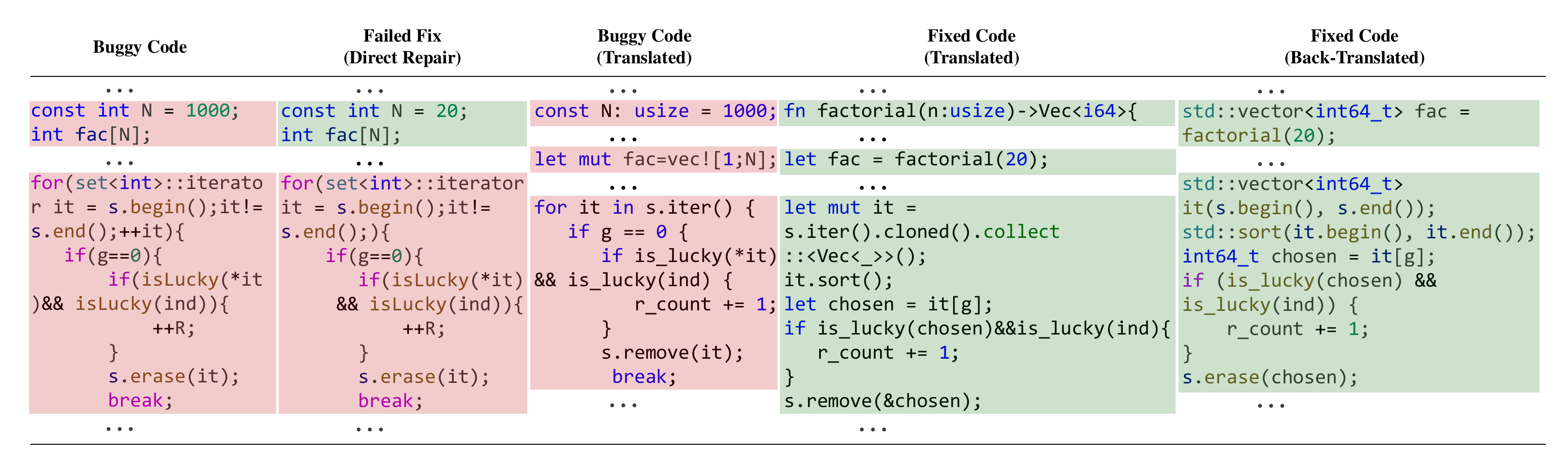}
  \caption{A motivating example of fixing a C++ bug by translating it to Rust.}
  \label{motivation-2}
  \vspace{-3mm}
\end{figure*}
Traditional rule-based translation approaches are costly, requiring substantial manual effort from programmers with specialized expertise in both the source and target languages \cite{ramos2024batfix}. Existing translation tools typically specialize in specific language pairs (e.g., Java to C\# \cite{chen2018tree}) and require significant domain-specific knowledge to develop and maintain \cite{bhatia2025verified}. Meanwhile, unsupervised neural approaches, while promising, necessitate high-quality data for each language pair and computational resources for model training, making them also resource-intensive \cite{roziere2020unsupervised}.

LLMs have shown promising capability in translating code between programming languages \cite{pan2024lost}. However, recent research highlights a significant gap between translating isolated code snippets and migrating entire software repositories, which involves complex dependencies, build configurations, and project-specific architectures \cite{wang2024repotransbench, zhang2025skeleton, ou2024repository, yan2023codetransocean}. To tackle this, the field has shifted towards repository-level translation, developing new benchmarks and neuro-symbolic methods that combine program analysis with LLMs to manage complexity \cite{ibrahimzada2025alphatrans, zhang2025scalable}. These studies show that while fully automated, correct translation of large projects remains a challenge, the generated code can significantly reduce manual effort. Since program repair is the main objective of our paper rather than translation, we adopt LLM-based code translation for our approach to explore whether APR can be enhanced with code translation.


\subsection{LLM-Based Program Repair Agent}
LLM-based agents employ LLMs as the cognitive core to perceive and act based on environmental feedback, working toward specific goals through four essential components: Planning, Memory, Perception, and Action \cite{liu2024large}. The planning component generates plans with reasoning strategies tailored to the task. The memory component maintains a record of historical data, while the perception component processes environmental feedback to facilitate more effective planning. The action component executes concrete steps based on decisions made by the planning component. Building upon this, existing program repair agents such as ChatRepair \cite{xia2024automated}, AutoCodeRover \cite{zhang2024autocoderover}, and RepairAgent \cite{bouzenia2024repairagent} typically follow a pipeline consisting of four steps in an iterative paradigm: (1) the agent first generates potential fixes for the buggy code, (2) these fixes undergo patch validation through compilation, execution, and automated testing autonomously, (3) repair feedback from validation step is carefully analyzed to inform the next iteration of fixing, and (4) the process repeats until a solution passes all validation criteria or reaches a predetermined iteration limit. Moreover, multi-agent architectures further extend this paradigm by decomposing complex tasks into specialized modules. FixAgent \cite{lee2024unified} leverages multiple agents dedicated to autonomous bug fixing and bug localization. AgentCoder \cite{huang2023agentcoder} integrates specialized agents responsible for code generation, test generation, and code execution. Similarly, ACFix \cite{zhang2024acfix} employs a configuration comprising a generator for proposing fixes and a validator tasked with ensuring their correctness. In our study, we leverage multiple LLM-based agents for program repair, code translation, and decision-making.

\section{Illustrative Example}
We illustrate how code translation can assist with repairing bugs that the LLM failed to fix with two real examples in our study. Figure \ref{motivation-1} and Figure \ref{motivation-2} present two examples demonstrating the efficiency of program repair with code translation. The examples show (1) the original buggy code, (2) the failed fix from direct repair, (3) the translated buggy code in a new programming language, (4) the fixed code in the target language, and (5) the back-translated fixed code. This detailed breakdown shows how the buggy code is first translated, then repaired in the target language, and finally translated back to its original language.

Figure \ref{motivation-1} demonstrates a Python bug causing a time limit error due to inefficient brute-force iteration. Direct repair attempts only partially address this by removing the prefix issue but retain the performance bottleneck. When translated to C++, the algorithm was redesigned to iterate over bit positions rather than the entire range, dramatically reducing complexity. The optimized solution is then successfully back-translated to Python. Technically, \textbf{C++’s built-in facilities such as bitset and its emphasis on low-level efficiency} not only expose latent inefficiencies in the original language but also foster an algorithmic design that is both conceptually elegant and computationally superior. 

\begin{figure*}[!t]
  \centering
  \includegraphics[width=.80\linewidth]{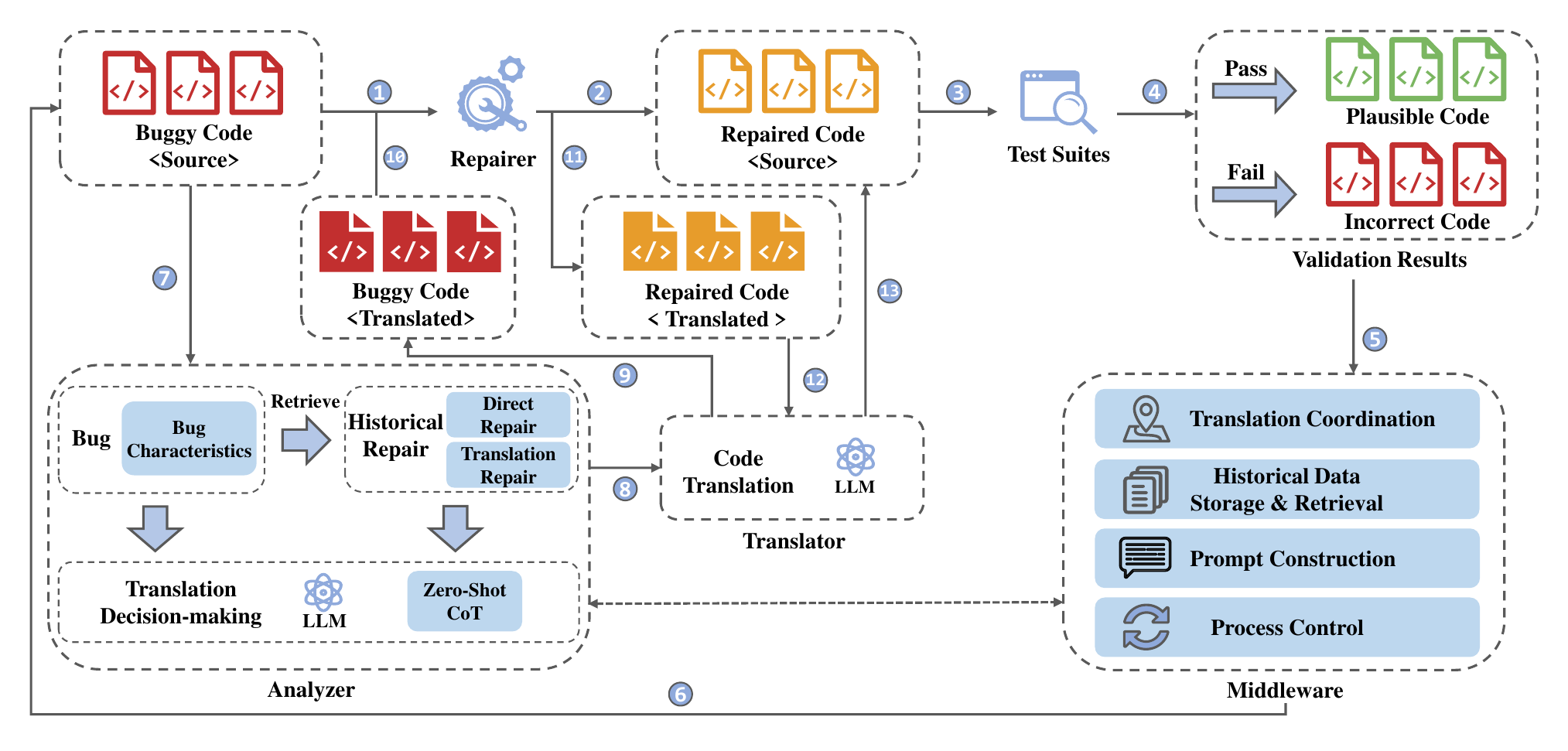}
  \caption{Overview of \lantern.}
  \vspace{-2mm}
  \label{overview}
\end{figure*}

Figure \ref{motivation-2} shows a C++ memory limit error. While direct repair failed to address the unsafe iterator usage, translating to Rust exposed the flaw immediately due to Rust's strict ownership rules (borrow checker). The Rust compiler's constraints forced a logic correction, which was preserved upon back-translation to C++, resolving the original memory limit issue. While the algorithm remains similar, \textbf{Rust's language-specific features regarding memory safety} help identify and fix memory and logical flaws that are less apparent in C++.

\section{Approach}
\subsection{Overview}
Figure \ref{overview} shows an overview of \lantern. The workflow is fully automated and begins with buggy code in its source programming language as input (step 1) to the repairer, where we employ LLM to fix bugs. The repairer generates the fixed code (step 2), which then undergoes validation against corresponding test suites of the bugs (step 3). The plausible code, the incorrect code, and their corresponding detailed evaluation results (step 4) are then forwarded to the middleware for further processing (step 5). The middleware identifies the unfixed bugs for subsequent translation-based repair (step 6) and directs them to the analyzer (step 7).
In addition to the bug characteristics, the analyzer retrieves historical records of fixing similar bugs through the middleware as historical feedback to inform its autonomous decision-making regarding the optimal target language for translation. Our approach constructs context-aware prompts for the reasoning process by dynamically integrating bug-specific details, historical fix insights from previous repair attempts.
Once the target language is decided after a comprehensive analysis by the LLM, the unfixed bugs are translated to their corresponding target languages (steps 8-9), and subsequent repair attempts are made for the translated bugs (steps 10-11). The repaired translated code is then returned to the translator (step 12), where it is back-translated to its source programming language (step 13), enabling the next round of validation and further refinement through subsequent iterations. To ensure an efficient exploration of attempts on language diversity, each target language is employed only once per bug within a single iteration in our study. 

\subsection{Middleware}
The middleware serves as the central coordination that orchestrates the key events of the entire workflow. As presented in Figure \ref{overview}, the middleware comprises four primary components as follows:

\textbf{Translation Coordination.} This component receives the evaluation outcomes, identifies the failed fixes, and schedules the corresponding unfixed bugs for subsequent translation-based repair.

\textbf{Historical Data Storage \& Retrieval.} Responsible for collecting and providing historical repair feedback for future reference. This component maintains a local vector database that encodes all evaluated bugs. The database is periodically updated after each iteration of repair. Historical repair feedback is sourced from two stages: (1) the initial direct repair attempts before translation-based repair, and (2) translation-based repair. While initial direct repair feedback is recorded at the beginning, translation-based repair feedback is updated in each iteration. Specifically, the evaluation results (e.g., repair performance, successful target languages where bugs are correctly fixed, etc.) and the bug characteristics (e.g., bug language, problem difficulty, execution outcome, etc.) from both sources of bugs are encoded into the vector database to facilitate effective bug retrieval.

\textbf{Prompt Construction.} This component dynamically constructs prompts for the LLM to support core actions in the workflow, including program repair, translation decision-making, code translation, and code back-translation according to specific bugs. In particular, it constructs prompts based on a designed prompt structure by incorporating detailed information such as bug characteristics along with auxiliary context like historical feedback from previous repair attempts, equipping the LLM with the necessary context to generate accurate responses for all tasks.

\begin{figure*}[!t]
  \centering
  \includegraphics[width=.80\linewidth]{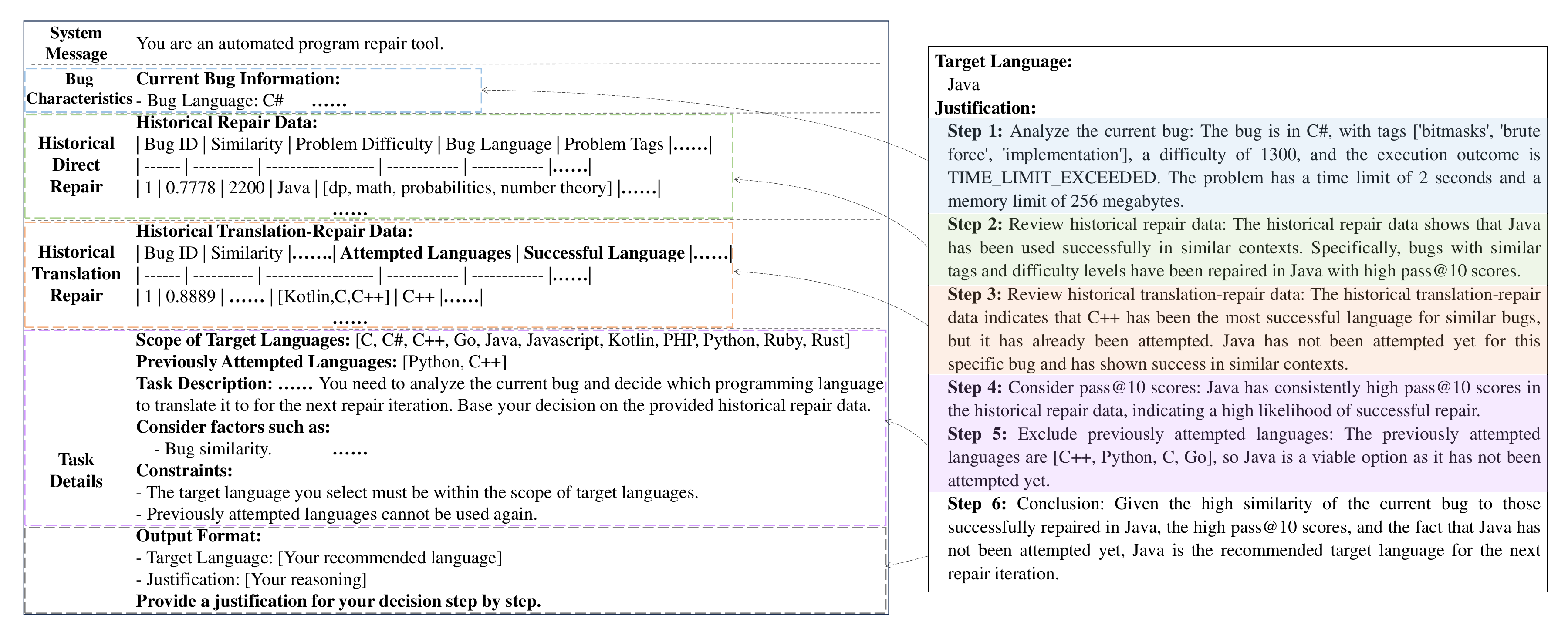}
  \caption{Prompt structure of translation decision-making (left) and an example of LLM response (right).}
  \label{decision-response}
\end{figure*}

\textbf{Process Control.} This component monitors the overall pipeline and terminates the workflow when predefined conditions are met (e.g., reaching a maximum iteration limit). In our study, the maximum iteration limit is designed based on the number of supported languages in the benchmark, allowing the system to explore the full range of potential translation paths, which in our experiments ranges from 1 to 11 iterations corresponding to the language scope.


Subsequently, any code that has been repaired in the target language then undergoes back-translation to its original programming language, thereby facilitating further refinements in iterative repair cycles.

\subsection{Analyzer \& Translator}\label{at}
As the most important components of the pipeline, the analyzer and translator manage the translation process of each bug including translation decision-making, translation, and back-translation. The LLM-based analyzer reasons about the environmental feedback and makes decisions autonomously. The translator is also built on an LLM foundation to enable translation across diverse languages. Once received, the bug is encoded as a bug query to retrieve similar bugs through the middleware, which returns the top-$k$ historical repair based on cosine similarity between the query bug and those from initial direct repair and previous translation-based repair. Consequently, the aggregated bug characteristics and the retrieved historical repair of similar bugs are provided as historical feedback for the analyzer. Specifically, the two sets of historical feedback serves a different purpose, respectively, where (1) the initial direct repair experience is provided for the analyzer to reason about which programming languages of bugs with similar characteristics the repairer can effectively resolve, and (2) the historical translation-based repair helps the LLM reveal optimal target languages previously employed in which similar bugs are successfully repaired. Once the optimal target language is determined, the bug is translated accordingly. 


\subsection{Prompt Design}

Our program repair prompt integrates a concise problem description, error type, input/output specifications, the buggy code, and a repair instruction. Similarly, our code translation prompt specifies the source and target languages, the corresponding code, and the translation instruction, with back-translation using the reversed source and target languages.

We also design the prompt dedicated to translation decision-making as illustrated in Figure \ref{decision-response} in a zero-shot chain of thought (CoT) \cite{jin2024zero} paradigm, which systematically integrates bug characteristics, historical repair feedback, and task constraints into a structured reasoning framework. Through this step-by-step process, the model autonomously identifies optimal target languages that demonstrate superior capabilities in fixing similar bugs.

\section{Evaluation}\label{evaluation}
\subsection{Research Questions}
We evaluate our approach on the following research questions:

\vspace{1mm}
\noindent\textbf{RQ1 (Overall Effectiveness):} 
\vspace{-1mm}
\begin{itemize}[leftmargin=*]
\item \textbf{RQ1.1 (Repair Performance)}: How effective is \lantern\ in enhancing program repair capabilities across different programming languages compared to direct repair? 
\item \textbf{RQ1.2 (Language Selection Efficiency)}: How efficiently does LLM-based reasoning in \lantern identify optimal target languages compared to other translation strategies?
\item \textbf{RQ1.3 (Approach Comparison):} How does the repair performance of LANTERN compare against other state-of-the-art (SOTA) program repair approaches?

\end{itemize}

\noindent\textbf{RQ2 (Translation Analysis):}
\begin{itemize}[leftmargin=*]
\item \textbf{RQ2.1 (Translation Outcomes)}: What underlying patterns behind the translation outcomes facilitate more efficient target language selection and enhance program repair?
\item \textbf{RQ2.2 (Translation Consistency)}: To what extent do the translation and back-translation process preserve semantic consistency between source and target languages?
\end{itemize}

\noindent\textbf{RQ3 (Ablation Study):} How do cross-language translation and historical feedback contribute to \lantern's repair performance and LLM-based reasoning efficiency in target language selection?

\noindent\textbf{RQ4 (Generalizability):}
\begin{itemize}[leftmargin=*]
\item \textbf{RQ4.1 (Real-World Generalizability):} How effectively does LANTERN generalize to real-world APR benchmarks?
\item \textbf{RQ4.2 (Model Generalizability):} Can LANTERN generalize to other LLMs and to what extent does LANTERN's effectiveness generalize across different LLMs?
\end{itemize}

\begin{figure*}[t]
  \centering
  \begin{subfigure}{0.33\linewidth}
    \centering
    \includegraphics[width=\linewidth]{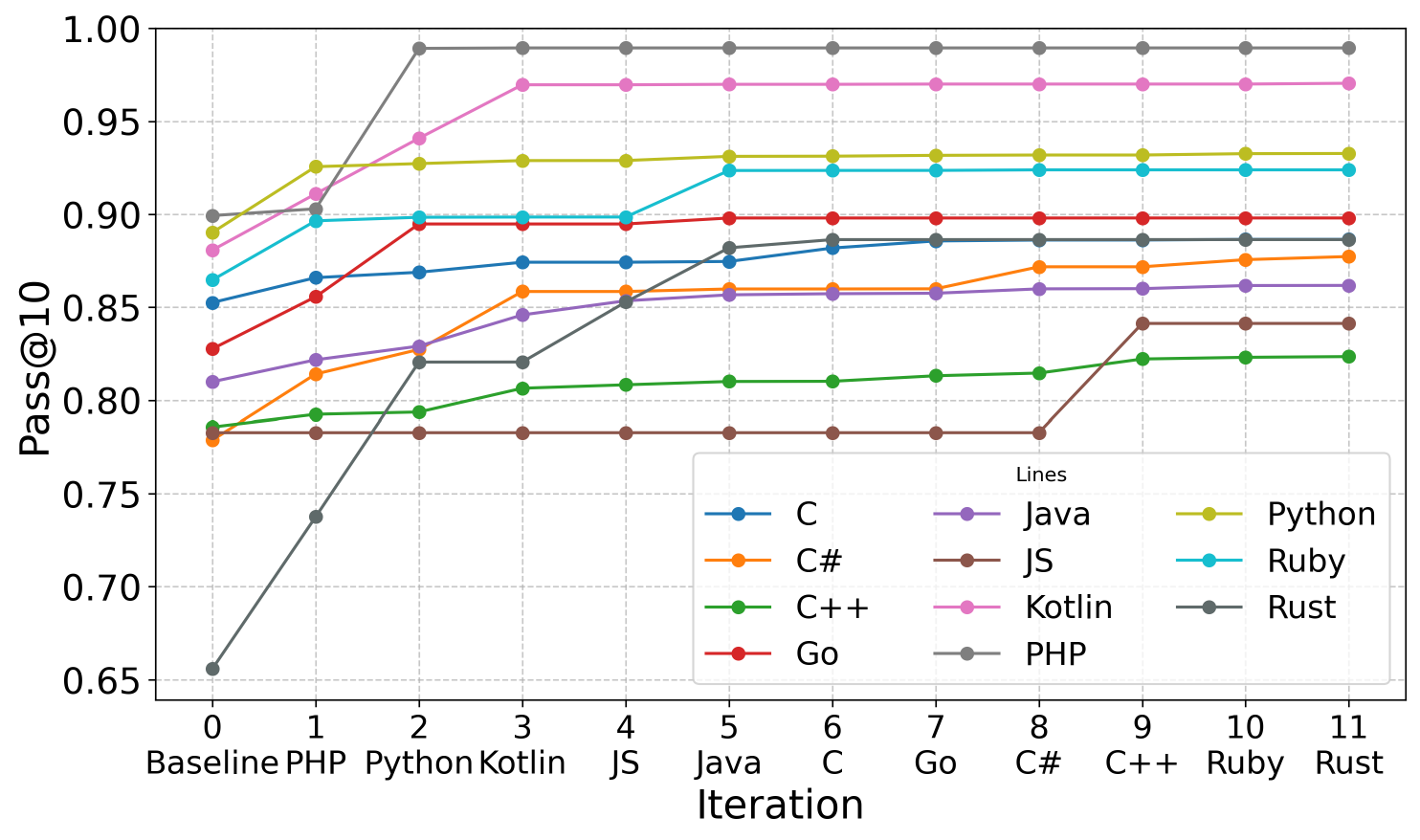}
    \caption{Greedy strategy (target languages on x-axis).}
    \label{tr_greedy}
  \end{subfigure}
  \begin{subfigure}{0.33\linewidth}
    \centering
    \includegraphics[width=\linewidth]{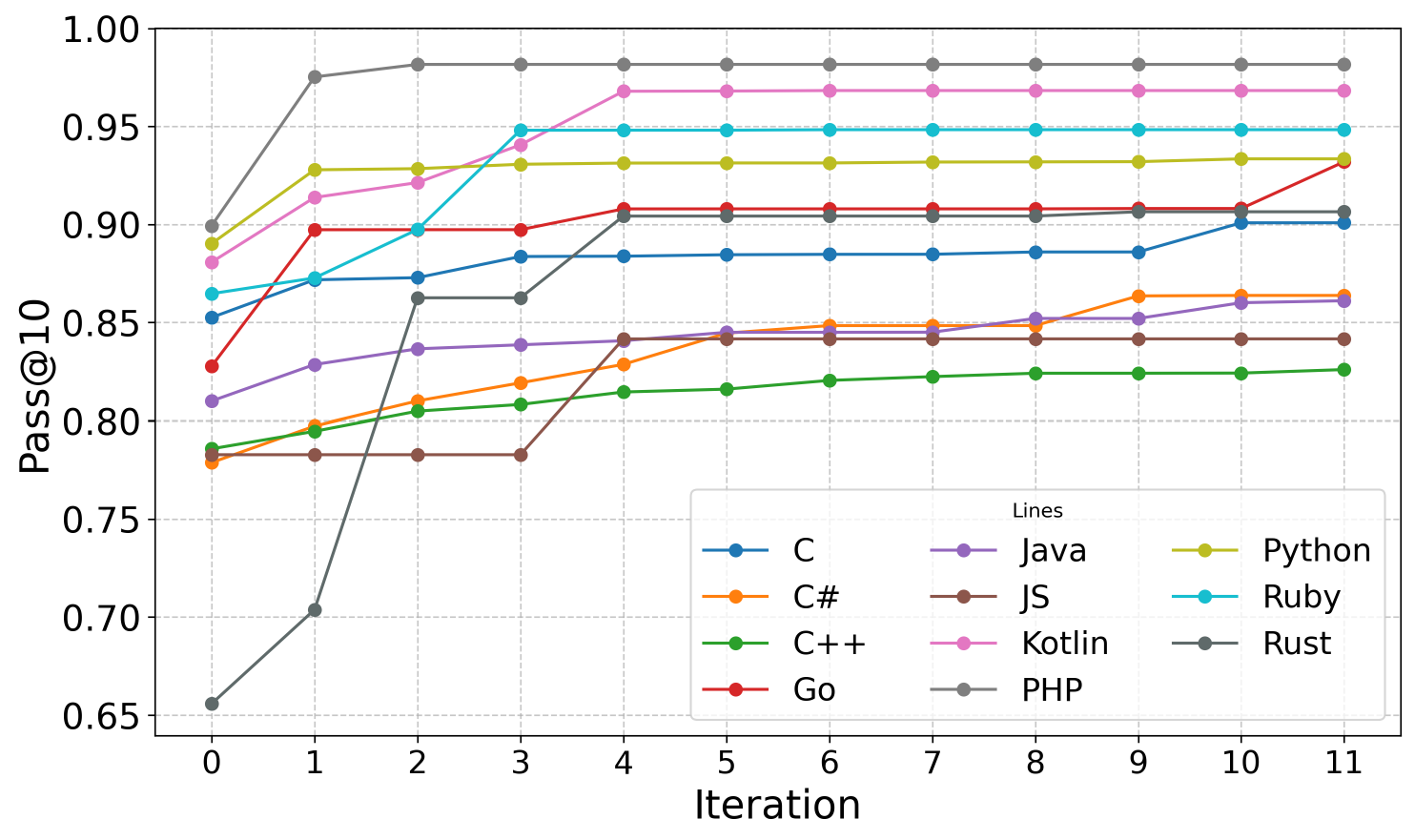}
    \caption{Random strategy.}
    \label{tr_random}
  \end{subfigure}
  \begin{subfigure}{0.33\linewidth}
    \centering
    \includegraphics[width=\textwidth]{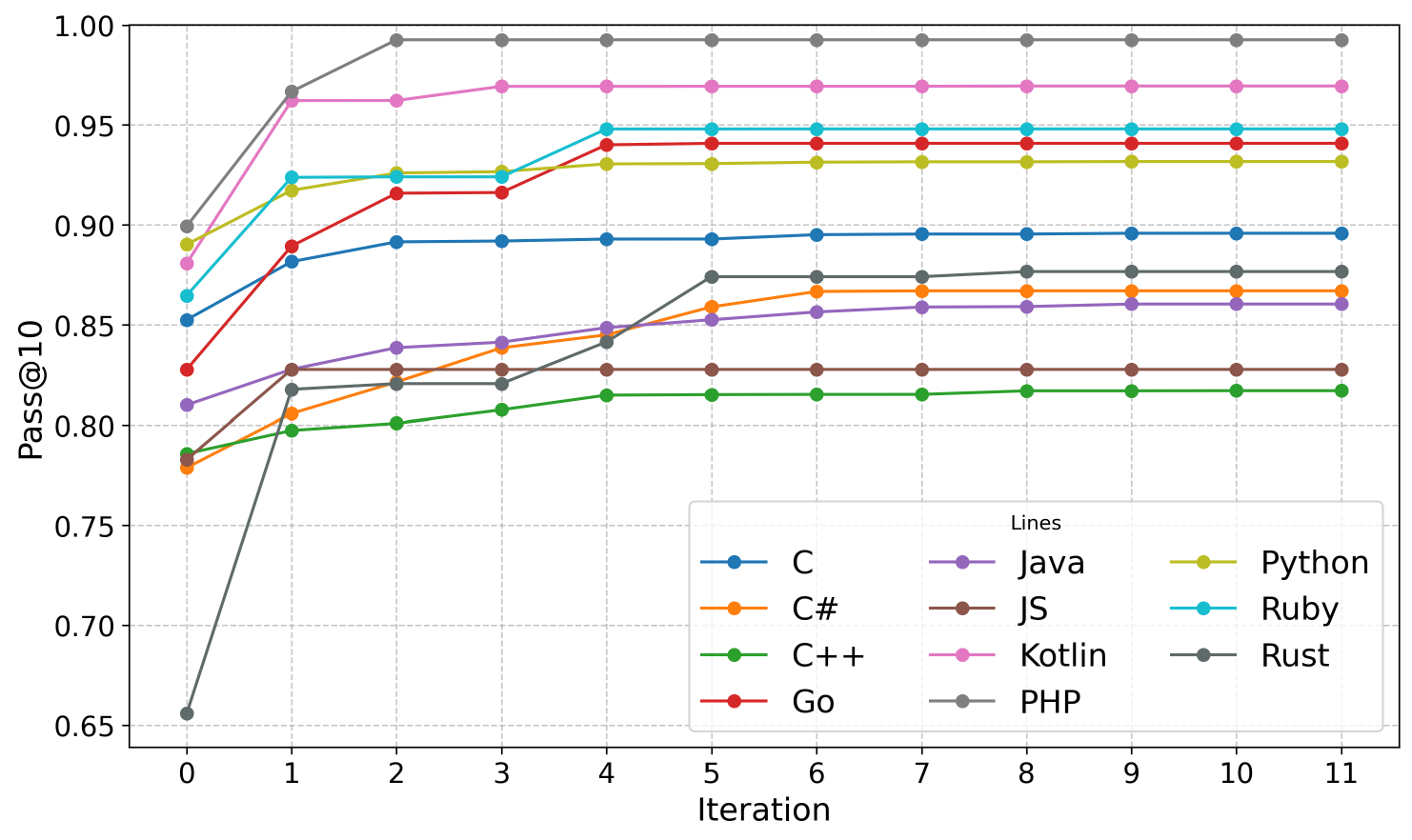}
    \caption{Reasoning strategy.}
    \label{tr_reasoning}
  \end{subfigure}
  \caption{Comparison of different strategies on Pass@10 across iterations (iteration 0 being the baseline of initial direct repair).}
  \label{strategies}
\end{figure*}

\subsection{Experiment Setup}\label{implementation}

\subsubsection{Model} 

For primary experiments from RQ1 to RQ3, we use the open-source state-of-the-art DeepSeek-V3 \cite{liu2024deepseek} model as the LLM in the analyzer and translator, for both decision-making and code translation. We also leverage the same LLM for the repairer to align the language scope with the translator instead of adopting various APR approaches specialized for different languages of bugs. We choose DeepSeek-V3 because it is arguably among the best for coding tasks in the literature.

To validate the generalizability of our approach, we include two additional prominent LLMs in RQ4. We selected one closed-source model, Claude 3.5 Sonnet \cite{claude3.5} and another powerful open-source model, Qwen2.5-72B-Instruct \cite{qwen2.5, qwen2}, as a SOTA open-source alternative. This selection allows us to assess whether our approach's benefits are tied to a specific model architecture or are more broadly applicable across different foundational models.

\subsubsection{Benchmark}\label{bm}
We employ three distinct and challenging benchmarks. To assess our approach from RQ1 to RQ3, we use the xCodeEval \cite{khan2024xcodeeval} benchmark. Specifically, we use the Compact Set, which is a subset dedicated for academic research, containing 5,068 bugs across 11 languages (Table \ref{dataset}). It is sourced from Codeforces \cite{codeforces} and covers difficulty ratings from 800 to 3,500. Unlike other alternatives such as HumanEval \cite{chen2021evaluating}, xCodeEval averages 50 tests per problem. In addition, xCodeEval's execution engine, ExecEval, provides dedicated runtime environments for all necessary compilers and interpreters, categorizing bug executions into six outcomes: compilation error, runtime error, memory limit exceeded, time limit exceeded, wrong answer, and passed.

For RQ4, we assess generalizability using Defects4J \cite{just2014defects4j} and SWE-Bench \cite{jimenez2023swe}. We utilize Defects4J v1.2 (391 bugs) and v2.0 (438 bugs) for Java projects. For repository-level Python repair, we use SWE-Bench Lite, comprising 300 real-world GitHub issues. This subset is notably difficult, with relatively low SOTA resolve rates, making it ideal for evaluating robustness.

\begin{table}[h]
\fontsize{7.5}{7.5}\selectfont
\centering
\caption{Dataset statistics of xCodeEval for each programming language.}
\label{dataset}
\begin{tblr}{
  width = \linewidth,
  colspec = {Q[63]Q[63]Q[75]Q[63]Q[71]Q[63]Q[94]Q[71]Q[112]Q[85]Q[75]Q[81]},
  hlines,
}
C   & C\# & C++ & Go  & Java & JS  & Kotlin & PHP & Python & Ruby & Rust & Total \\
733 & 739 & 641 & 294 & 716  & 183 & 313    & 191 & 710    & 343  & 205  & 5068  
\end{tblr}
\end{table}

\subsubsection{Metrics}
We employ evaluation metrics from multiple dimensions to rigorously assess the results in both program repair and decision-making.

\textbf{\textit{Pass@k}} measures the likelihood that at least one of the top $k$ generated patches fixes a bug \cite{chen2021evaluating,zheng2023codegeex}.
Since developers are generally willing to review a maximum of approximately 10 patches~\cite{noller2022trust}, we use $Pass@10$ under $n=20$.

We also evaluate the reasoning ability of our approach, where the selected optimal target languages form a ranked list of languages over iterative attempts until the bug is fixed. Similar to recommendation systems that evaluate the rankings of \textbf{relevant} items, we assess our ranking quality based on \textbf{valid} iterations (i.e., where the selected languages result in successful fixes). We employ the ranking metrics \cite{jadon2024comprehensive} as follows:

\textbf{\textit{Precision@k}} quantifies the proportion of valid iterations in the top $k$ iterations. 

\textbf{\textit{Mean Average Precision at k (MAP@k)}} evaluates the overall quality of the rankings of target languages.


\textbf{\textit{Normalized discounted cumulative gain at k (NDCG@k)}} measures the ranking quality by giving higher weights to valid iterations that appear earlier.%

\textbf{\textit{Recall@k}} measures the proportion of valid iterations among the top $k$ iterations in the number of all relevant iterations. 

\textbf{\textit{F1@k}} is the harmonic mean of $Precision@k$ and $Recall@k$, providing a balanced view between the two metrics.

\subsection{RQ1: Overall Effectiveness}
\subsubsection{Experimental Design}

To evaluate the effectiveness of our approach, we establish a maximum of 11 iterative cycles since there are 11 available programming languages in the benchmark, ensuring that all different target languages are attempted for each bug. In contrast to picking any language for each bug at each iteration, our approach leverages an LLM-based decision-making strategy that identifies the optimal target language by reasoning about the bug characteristics and historical feedback from previous repair attempts. Considering that all fixable bugs can always encounter a successful target language by the end of all 11 iterations, we compare with two more strategies as follows:
\begin{itemize}[noitemsep,topsep=0pt,leftmargin=1em]
    \item \textbf{Greedy Strategy:} In this strategy, target languages are prioritized based on the historical performance of the LLM in the initial repair stage. For example, if the LLM achieved the best $Pass@10$ score on PHP initially, then PHP is selected as the target language in the first iteration for all bugs.
    \item \textbf{Random Strategy:} A target language is randomly selected for each bug in each iteration, serving as another baseline for evaluating the efficiency of the LLM-based decision.
    \item \textbf{Reasoning Strategy:} The LLM-based decision-making strategy that reasons about the optimal target language based on bug characteristics and historical feedback.
\end{itemize}
We compare the $Pass@10$ scores across iterations for all strategies to evaluate the repair performance in RQ1.1. We evaluate the quality of target language selection in RQ1.2 using ranking metrics that assess each strategy's efficiency in identifying valid repairs, which is critical since resolving more bugs in earlier iterations substantially reduces subsequent computational costs.

In RQ1.3, we compare LANTERN against SOTA approaches using DeepSeek-V3 on xCodeEval. Baselines include direct repair (original xCodeEval prompting), ChatRepair (iterative conversational refinement), Self-Planning (high-level algorithmic planning before coding), and Self-Collaboration (virtual development team with different agent roles for analysis, fixing, and validation).

\begin{figure}[h]
  \centering
  \includegraphics[width=.8\linewidth]{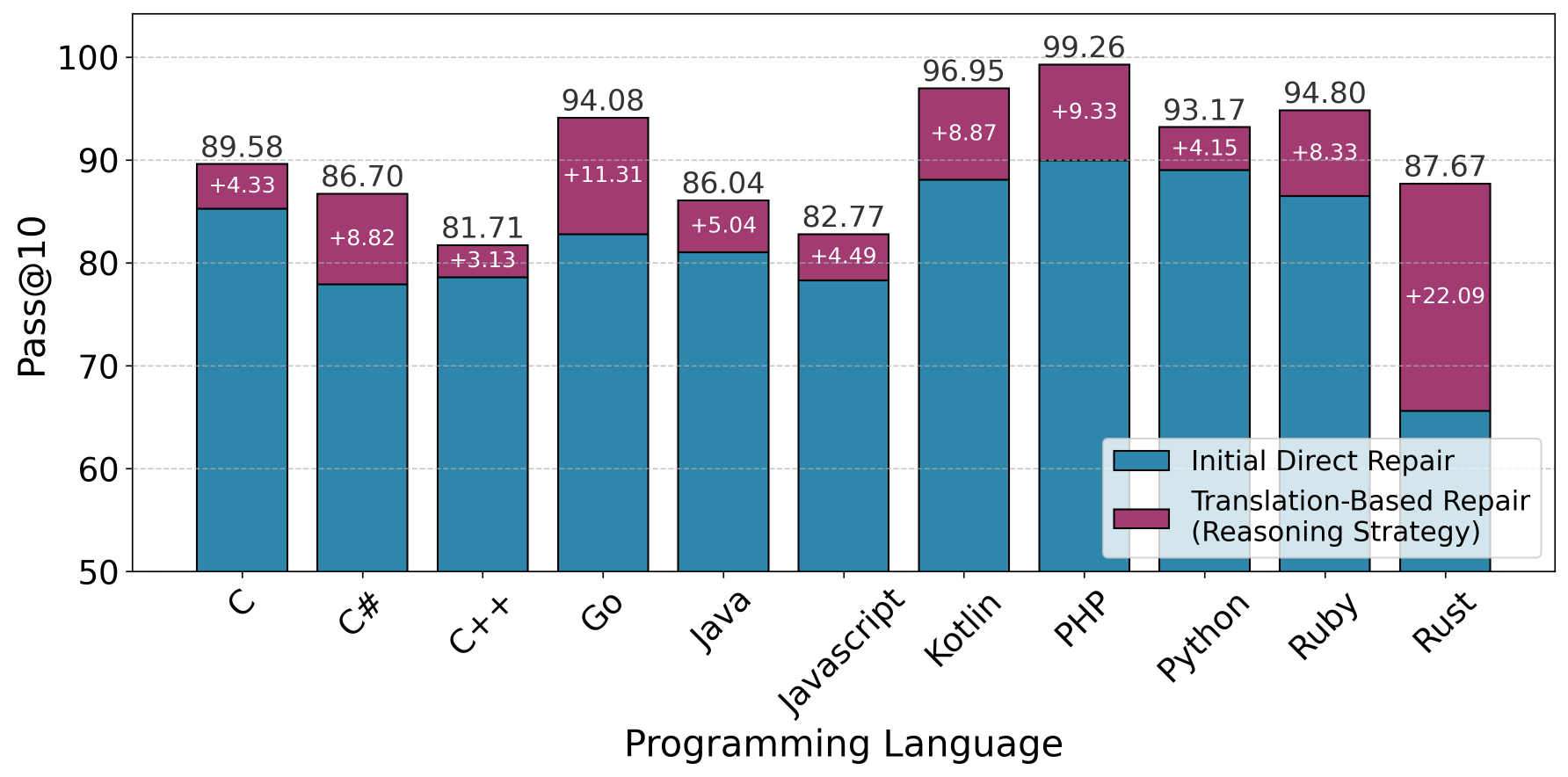}
  \caption{Pass@10 improvements of the reasoning strategy across programming languages.}
  \label{pass10_imp}
\end{figure}

\subsubsection{Experimental Results for RQ1.1 (Repair Performance)}
Figure \ref{strategies} shows the $Pass@10$ scores of the three strategies across iterations for bugs from 11 programming languages. 
We first observe in Figure \ref{tr_greedy} that under the greedy strategy, some languages require several iterations to yield fixes. For example, JavaScript (JS) bugs are only repaired in the ninth iteration after translation to C++, and PHP bugs cannot be fixed until the second iteration. 
\begin{figure}[h]
  \centering
  \includegraphics[width=.8\linewidth]{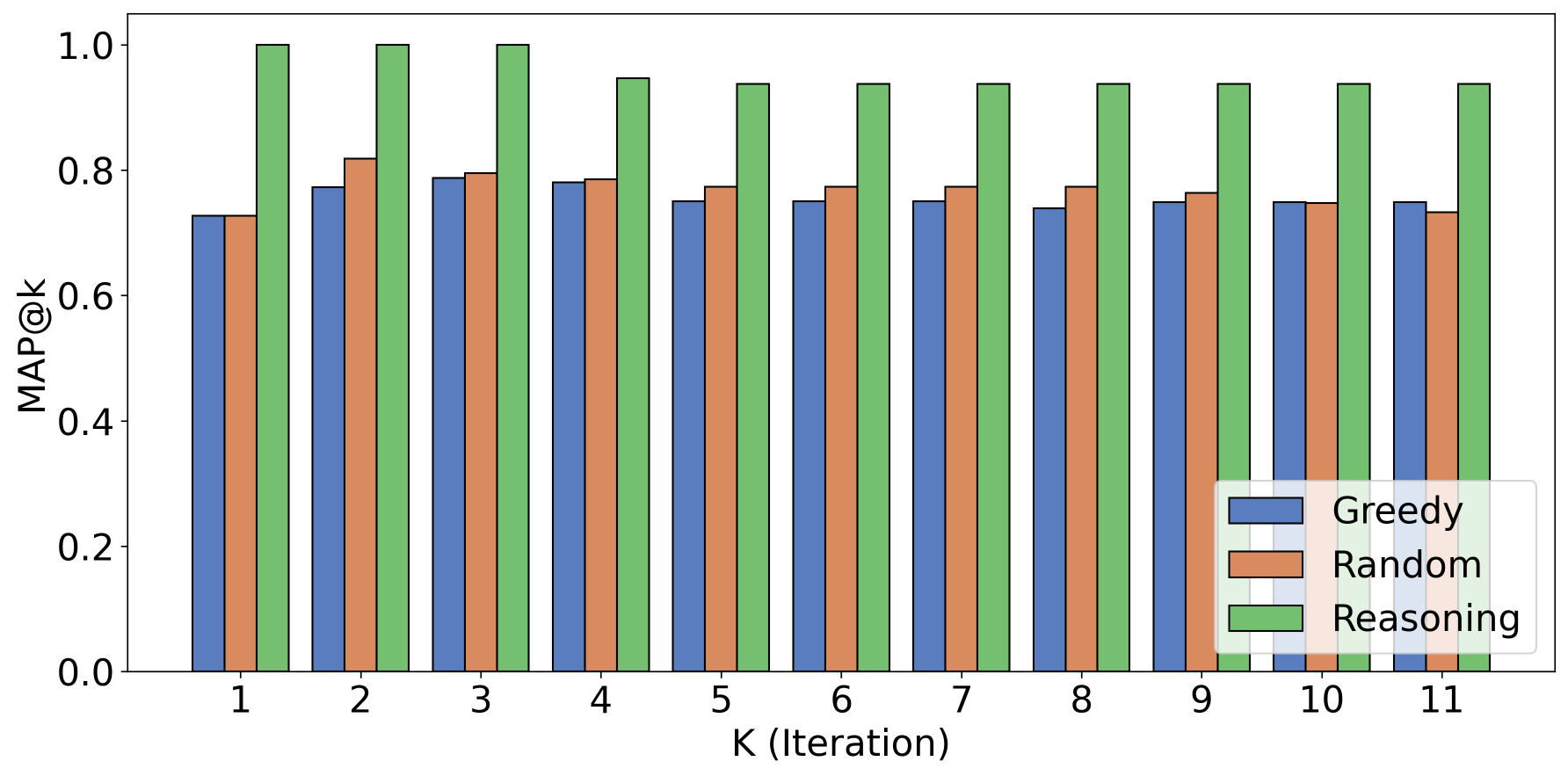}
  \caption{MAP@k of different strategies across iterations.}
  \label{map}
\end{figure}
In contrast, the random strategy produces a uniform distribution of target languages that leads to earlier fixes, where most PHP bugs are fixed immediately, with Go and Rust bugs achieving $Pass@10$ scores of 0.90 and 0.86 in the first and second iterations, respectively. However, JS bugs still cannot be fixed until the fourth iteration. Notably, the reasoning strategy as illustrated in Figure \ref{tr_reasoning} fixes most bugs in the first iteration, with JS bugs resolved immediately and Kotlin and Ruby reaching $Pass@10$ scores of 0.96 and 0.92 in the first iteration, indicating the superior ability of LLM-based decision-making to select optimal target languages early on.

On the other hand, Figure \ref{pass10_imp} presents the final $Pass@10$ improvements of the reasoning strategy on all languages. It is worth noting that Rust bugs benefit the most from translation-based repair, which achieves an increment of 22.09\%. Following are the Go, PHP, Kotlin, C\#, and Ruby bugs that undergo improvements of 11.31\%, 9.33\%, 8.87\%, 8.82\%, and 8.33\%. Our MWW tests \cite{mann1947test,wilcoxon1992individual} further confirm the improvements with a p-value of 0.0126 ($<0.05$) and a Cliff's Delta \cite{cliff1993dominance,yang2024federated} of 0.64 (large effect).

\begin{figure}[h]
  \centering
  \includegraphics[width=.8\linewidth]{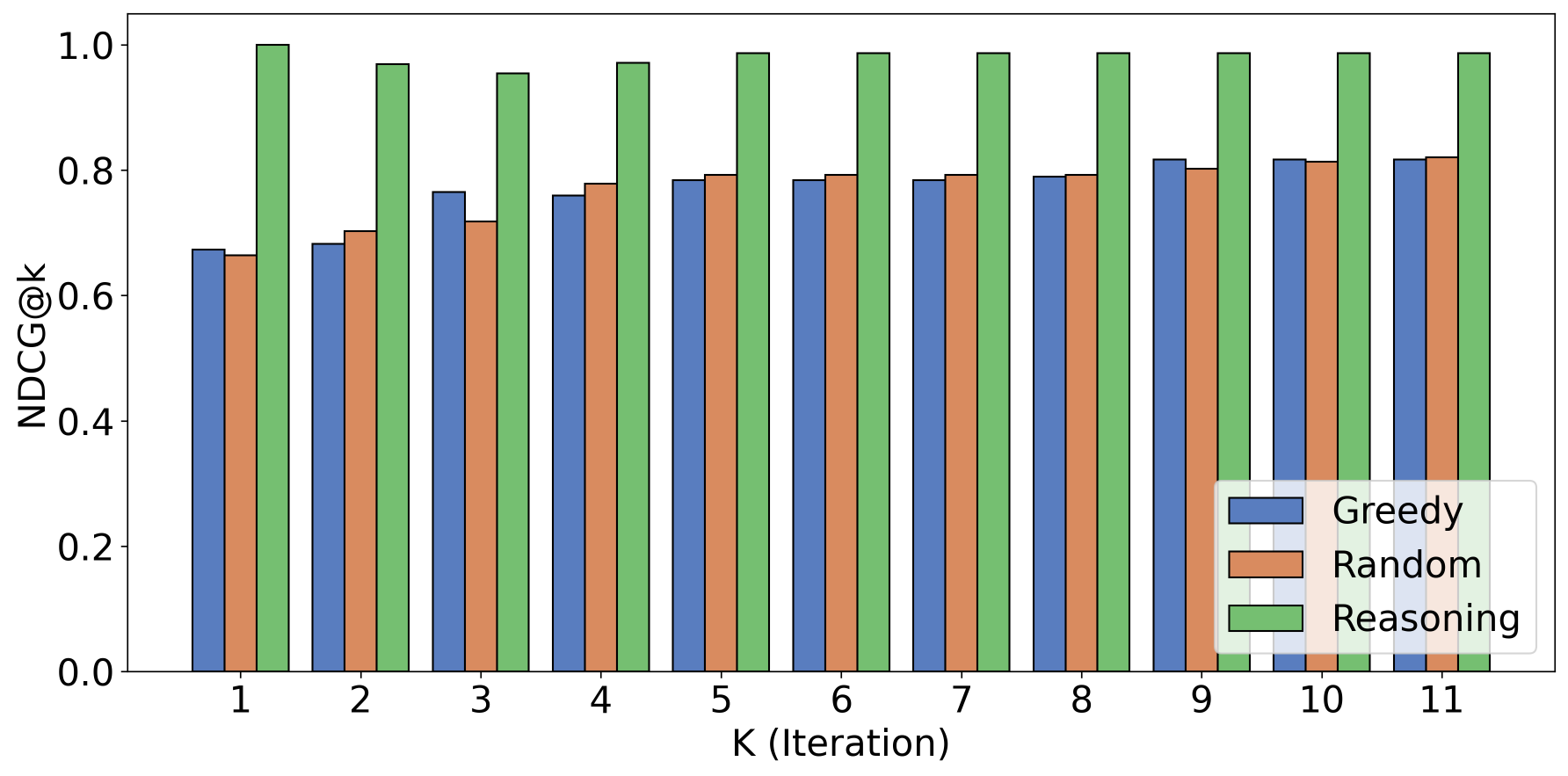}
  \caption{NDCG@k of different strategies across iterations.}
  \label{ndcg}
\end{figure}

\subsubsection{Experimental Results for RQ1.2 (Language Selection Efficiency)}
Figure \ref{map} shows that the reasoning strategy maintains an $MAP@k$ of 1.00 for the first three iterations and converges to 0.938 by the fifth iteration, reflecting its consistent ability to identify optimal languages. Figure \ref{ndcg} exhibits that the reasoning strategy starts with a $NDCG@k$ of 1.00 at the first iteration and consistently performs better than the other strategies at each iteration, highlighting its early advantages in the ranking. 

\begin{figure*}[h]
  \centering
  \begin{subfigure}{0.28\linewidth}
    \centering
    \includegraphics[width=\linewidth]{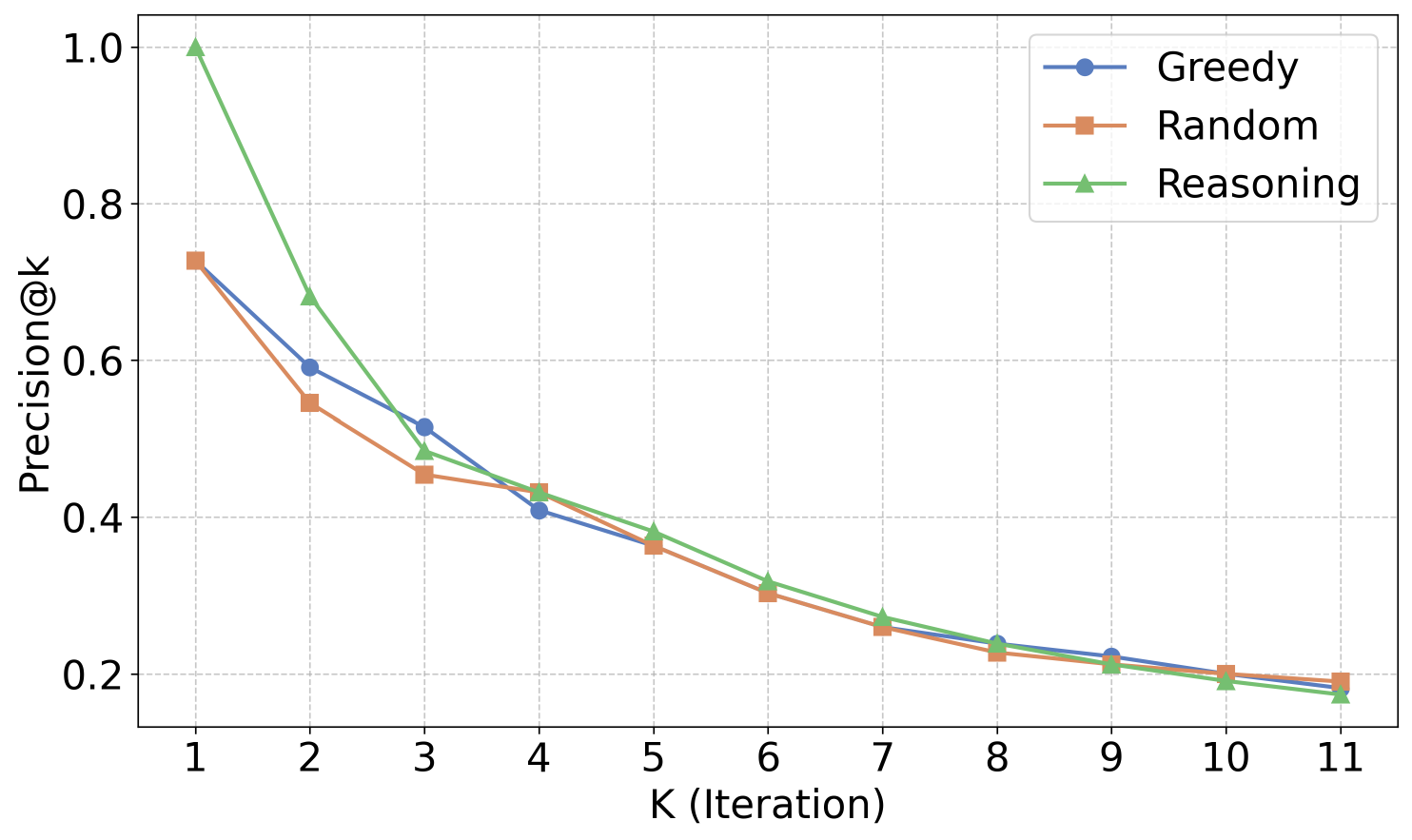}
    \caption{Precision@k.}
    \label{precision}
  \end{subfigure}
  \begin{subfigure}{0.28\linewidth}
    \centering
    \includegraphics[width=\linewidth]{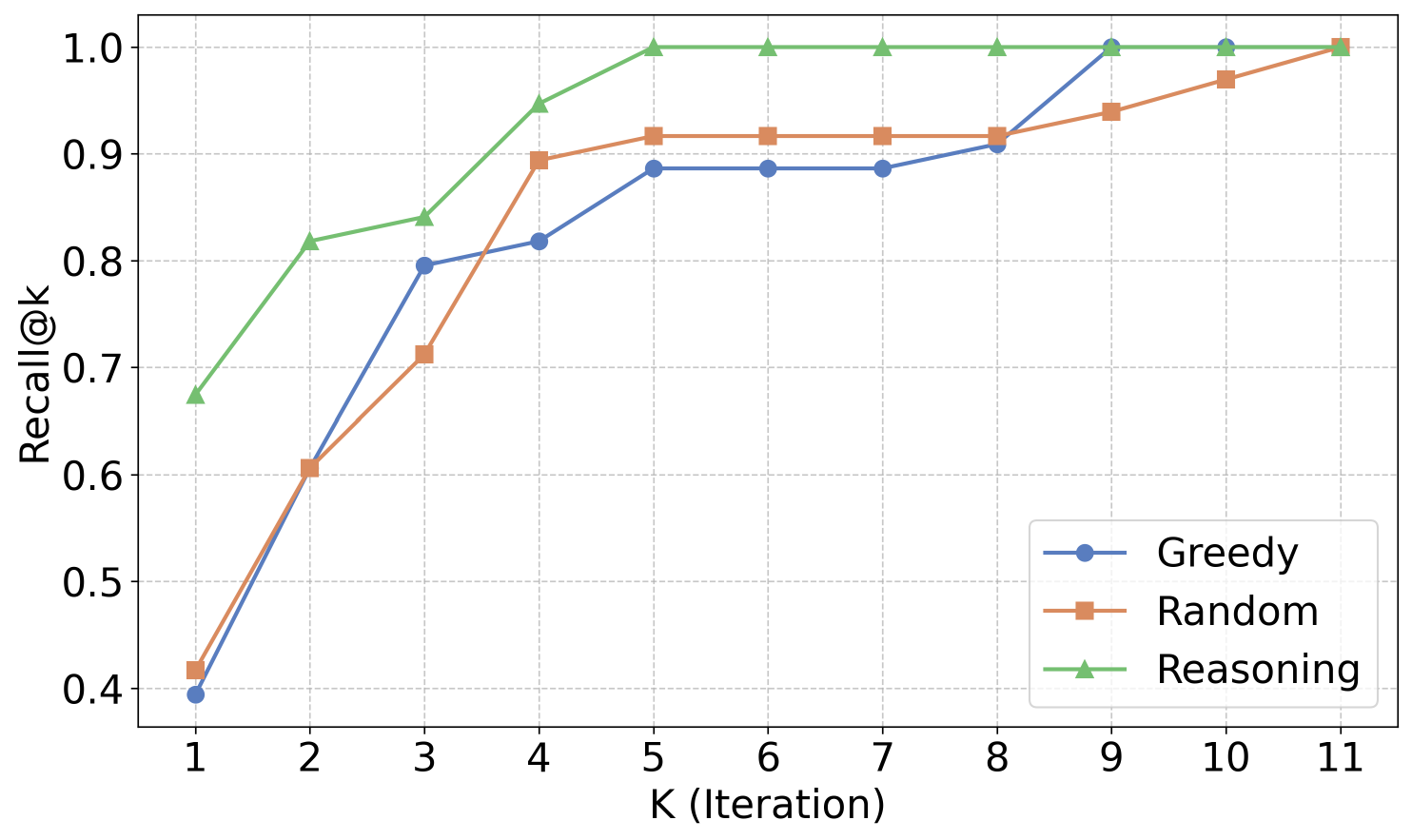}
    \caption{Recall@k.}
    \label{recall}
  \end{subfigure}
  \begin{subfigure}{0.28\linewidth}
    \centering
    \includegraphics[width=\textwidth]{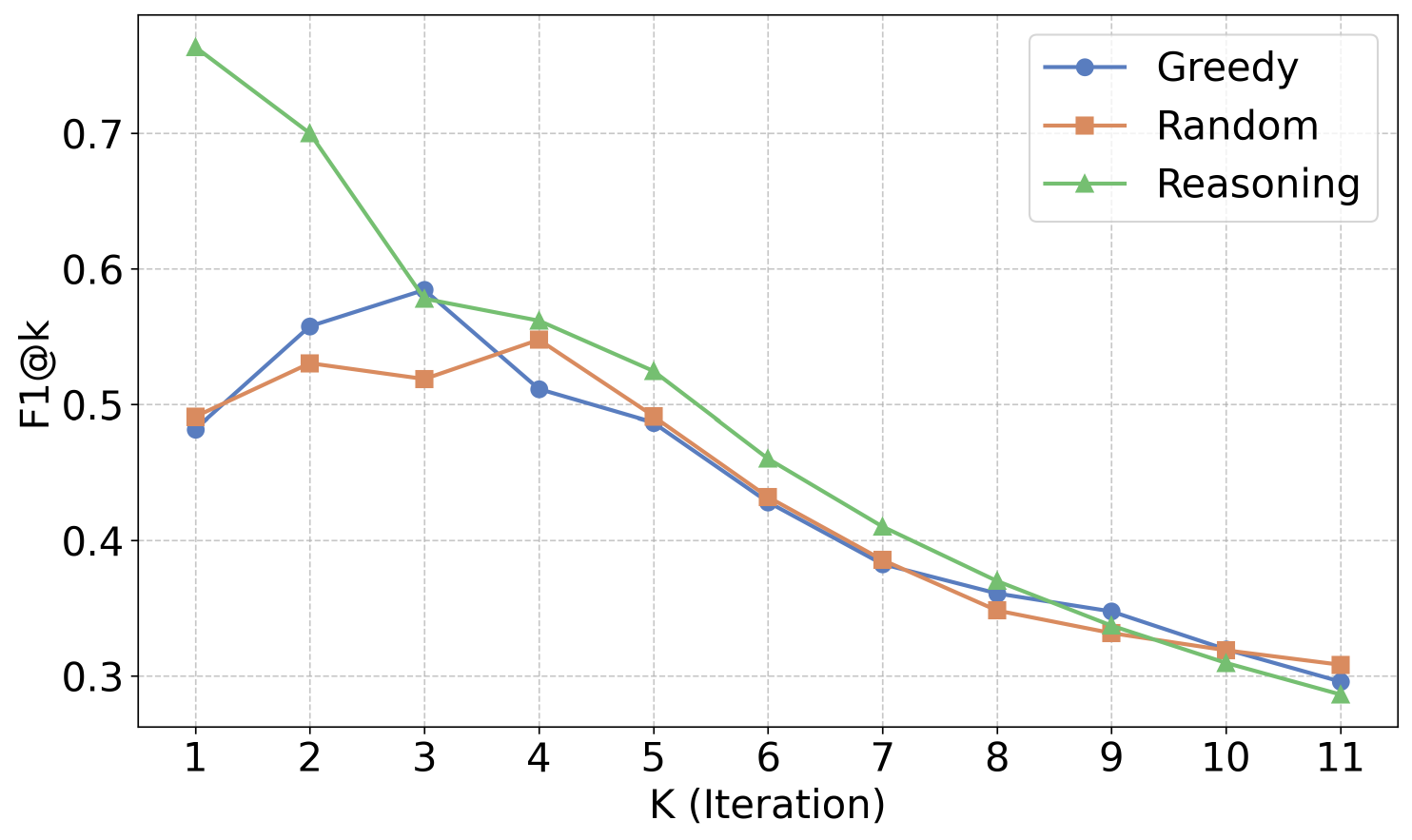}
    \caption{F1@k.}
    \label{f1}
  \end{subfigure}
  \caption{Comparison of different strategies on Precision@k, Recall@k, and F1@k across iterations.}
  \label{prf}
\end{figure*}


According to Figure \ref{precision}, the reasoning strategy remains higher $Precision@k$ until all strategies converge to the same level at the third iteration, demonstrating its early identification of valid fixes. Figure \ref{recall} shows that it achieves full recall with a $Recall@k$ of 1.00 by the fifth iteration, ahead of greedy and random strategies, which converge only at the ninth and final iterations, respectively. Similarly, the $F1@k$ scores in Figure \ref{f1} highlight its early advantage, aligning with our previous $Pass@10$ observations.

\subsubsection{Experimental Results for RQ1.3 (Approach Comparison)}
\begin{figure}[t]
  \centering
  \includegraphics[width=.9\linewidth]{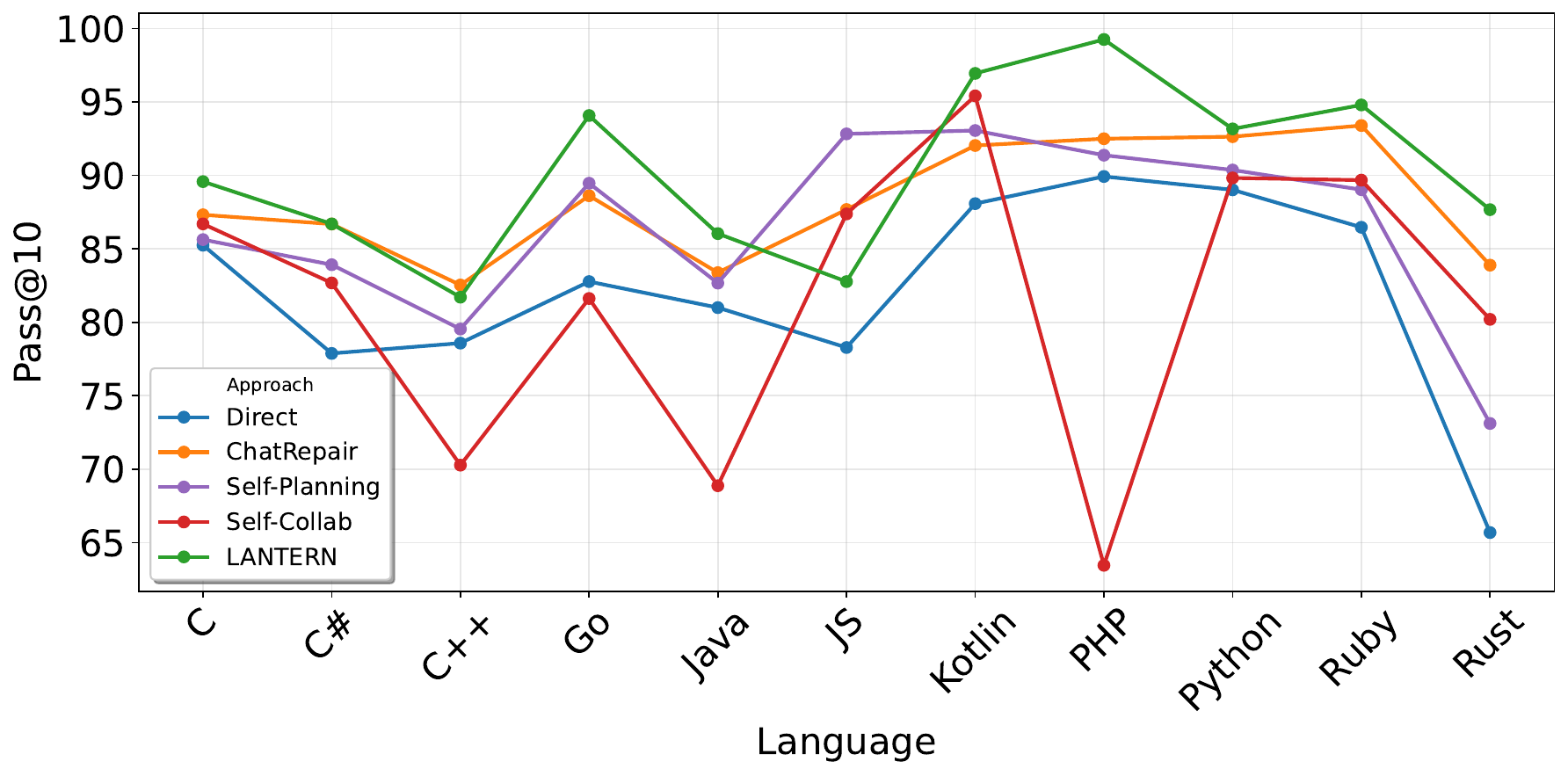}
  \caption{Pass@10 Performance of Different Approaches on xCodeEval}
  \label{rq1.3}
\end{figure}

Figure \ref{rq1.3} shows LANTERN achieves the highest $Pass@10$ scores across xCodeEval, leading in 9 out of 11 languages. ChatRepair serves as a strong iterative baseline, demonstrating considerable gains from refinement within the original language (e.g., +18.31\% for Rust). However, LANTERN consistently surpasses ChatRepair with improvements of 5.46\%, 4.91\%, and 6.76\% on Go, Kotlin, and PHP respectively, suggesting that repetitive refinement has performance ceilings when confined to a single language. This gap stems from key architectural differences: ChatRepair iteratively refines solutions within one conversation, risking context accumulation and potential loss of critical early information, while LANTERN treats each repair as an independent session, ensuring context control and achieving one-shot solutions.

Self-Planning shows promise in some languages due to high-level planning abilities, achieving notable gains in JavaScript (+14.55\%), but modest improvements elsewhere (e.g., +0.95\% in C++) and significantly underperforming LANTERN. This indicates that generating high-level plans is insufficient when the LLM's underlying implementation ability in specific languages is weak.

Self-Collaboration achieves large increments of 9.10\% and 14.62\% on JavaScript and Rust but causes severe performance degradation in several languages: C++ (-8.31\%), Java (-12.13\%), and most notably PHP (-26.48\%). This suggests overly complex, simulated multi-role interactions can be inefficient, with the system potentially hallucinating incorrect analyses or validations that further mislead the model.

\findx{\textbf{[RQ-1] Findings:} (1) Simply defaulting to the language with the model's historically best performance does not guarantee the most effective repairs. Instead, a more tailored approach that analyzes specific bugs to select the optimal target language leads to more robust repair performance. (2) \lantern can significantly enhance program repair, particularly for less common languages like Rust, achieving a maximal improvement of 22.09\% in Pass@10. (3) The LLM reasoning capability can significantly enhance translation decision-making, with an average MAP@k of 0.957 and NDCG@k of 0.954, enabling efficient target language selection. (4) Changing the linguistic context of a bug is a more powerful enhancement technique than simply iterating or reasoning more deeply within the original language, especially for complex bugs or in languages where the LLM's native proficiency is limited.}

\subsection{RQ2: Translation Analysis}
\subsubsection{Experimental Design}
In RQ2.1, we investigate the translation outcomes to reveal the underlying differences between strategies. We compare the translation paths of the reasoning strategy against the greedy and random strategies, and analyze bugs fixed via cross-language translation versus direct repair. In particular, we assess the code difficulty of bugs fixed by \lantern\ to determine if translation better addresses complex buggy code. Additionally, due to the partial semantic loss potentially caused by translation \cite{pan2024lost,ramos2024batfix}, we validate semantic consistency in RQ2.2 for both bug translation and code back-translation by comparing test outcomes before and after translation.

\begin{figure}[t]
  \centering
  \begin{subfigure}{\linewidth}
    \centering
    \includegraphics[width=\linewidth]{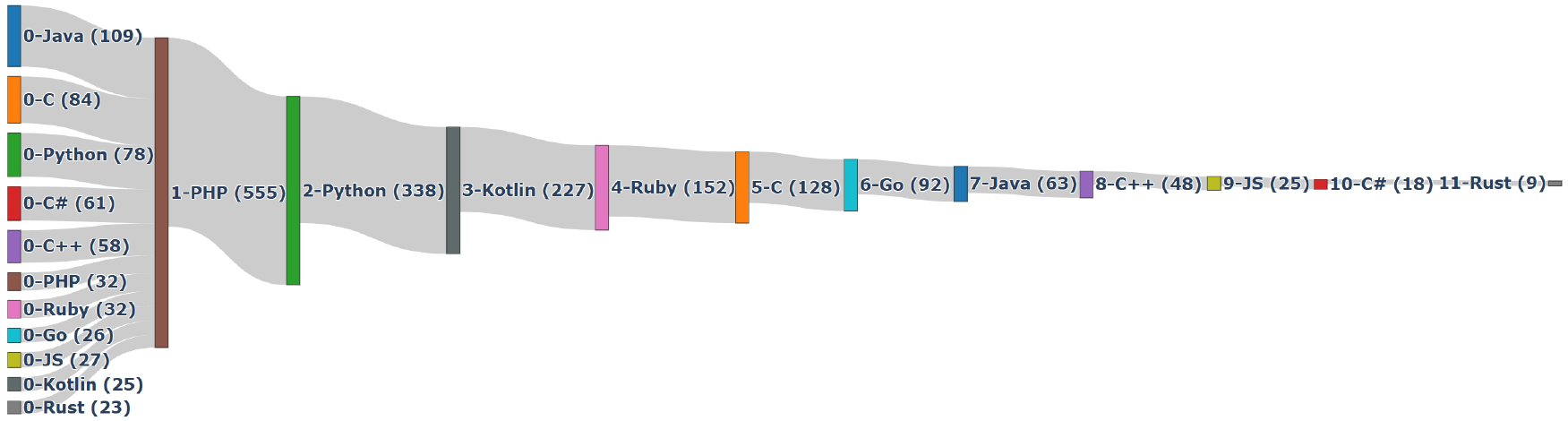}
    \caption{Translation paths of the greedy strategy.}
    \label{sk_greedy}
  \end{subfigure}
  \begin{subfigure}{\linewidth}
    \centering
    \includegraphics[width=\linewidth]{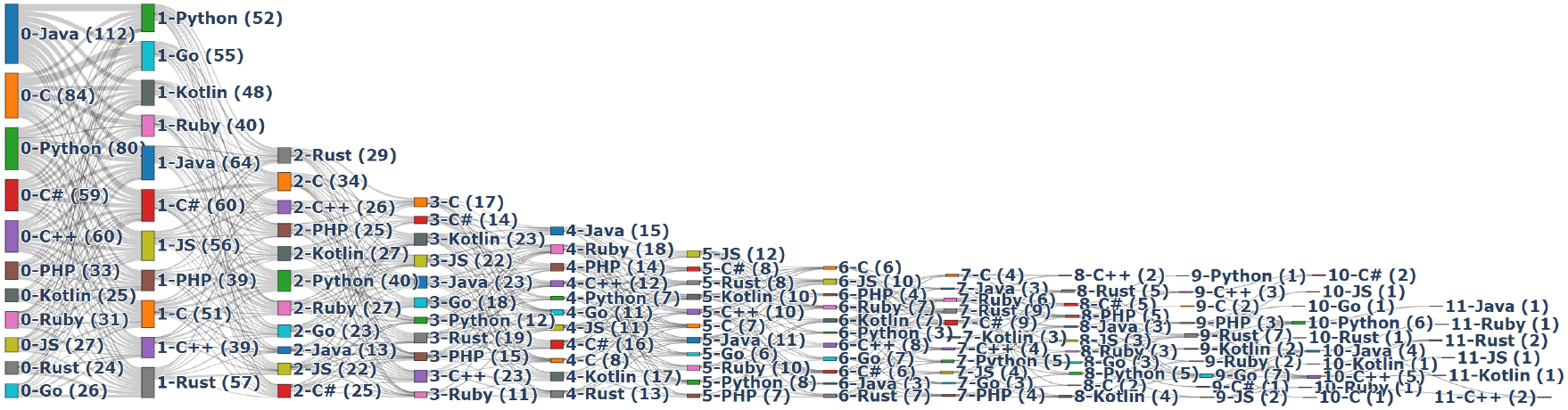}
    \caption{Translation paths of the random strategy.}
    \label{sk_random}
  \end{subfigure}
  \begin{subfigure}{\linewidth}
    \centering
    \includegraphics[width=\textwidth]{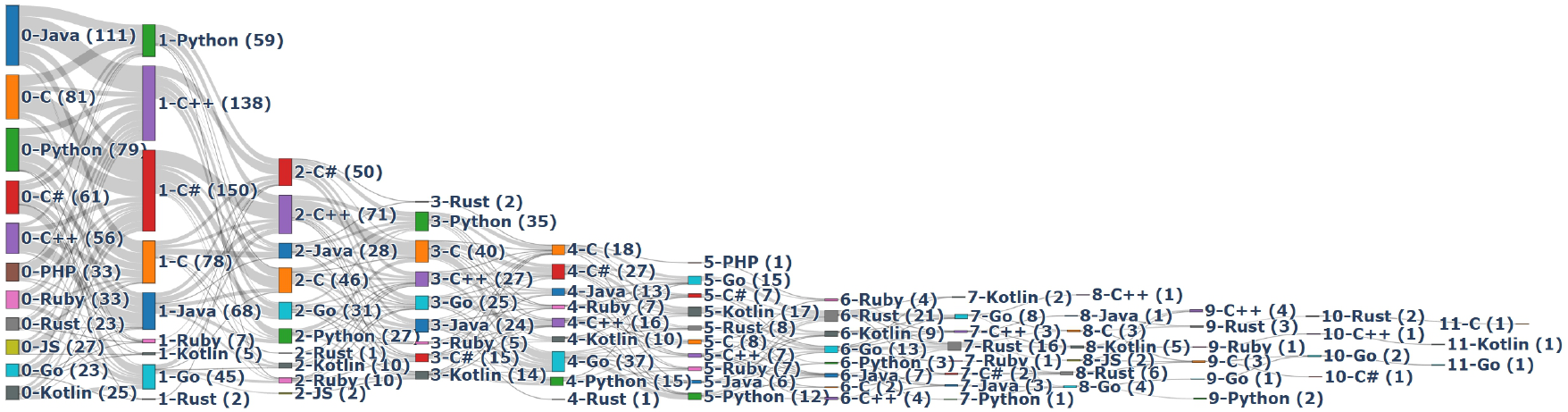}
    \caption{Translation paths of the reasoning strategy.}
    \label{sk_reasoning}
  \end{subfigure}
  \caption{Sankey diagrams of translation paths for successfully fixed bugs.}
  \label{sk}
\end{figure}

\subsubsection{Experimental Results for RQ2.1 (Translation Outcomes)}
Figure \ref{sk} shows the translation paths across iterations, with each node in the Sankey diagrams labeled as "\textit{iteration number - bug language (number of bugs)}", where iteration 0 is the initial phase. As illustrated in Figure \ref{sk_greedy}, all bugs are translated to a single language per iteration by the greedy strategy, while the random strategy exhibited in Figure \ref{sk_random} selects nearly equal target languages each time. In contrast, Figure \ref{sk_reasoning} demonstrates that the reasoning strategy gravitates towards specific languages. For example, the top-5 target languages in the first iteration, C\#, C++, C, Java, and Python account for 150, 138, 78, 68, and 59 bugs, respectively. Similar outcomes can be observed particularly in the first three iterations after which the selection begins to shift toward the remaining languages since most languages have already been attempted. Our evaluation demonstrates that the reasoning strategy achieves the shortest average translation path length (2.52), outperforming both random (2.69) and greedy (2.98) strategies, thereby minimizing resource consumption throughout the iterative process as unfixed bugs must propagate through all subsequent iterations.

\begin{figure}[t]
  \centering
  \includegraphics[width=.8\linewidth]{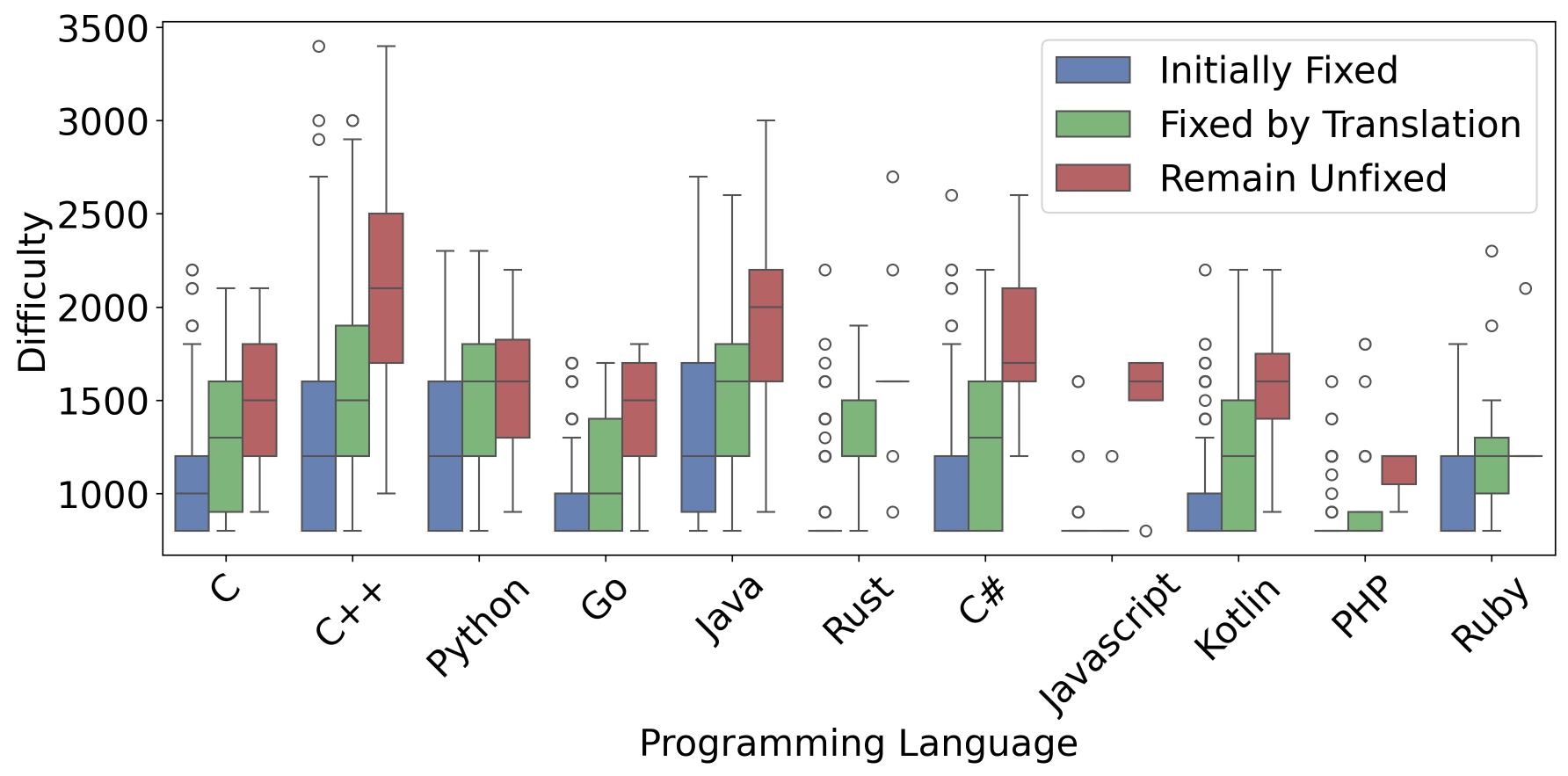}
  \caption{Difficulty distribution of the bugs fixed by initial repair, translation-based repair, and unfixed bugs.}
  \label{difficulty}
\vspace{-3mm}
\end{figure}

Figure \ref{difficulty} exhibits the difficulty distribution of the bugs initially fixed by direct repair, fixed via translation, and those remain unfixed. We notice that the translation-based repair can fix bugs with 500 higher median difficulty for C\#, while for Java, Kotlin, Ruby, and Rust, it enables successful repair of a group of bugs with median difficulty increased by 400. Our statistical analysis reveals a significant difference in difficulty between bugs fixed via translation and those initially fixed with a p-value of $5.25E-48$ and Cliff's delta of 0.37 (medium effect).

\begin{figure}[h]
  \centering
  \includegraphics[width=.85\linewidth]{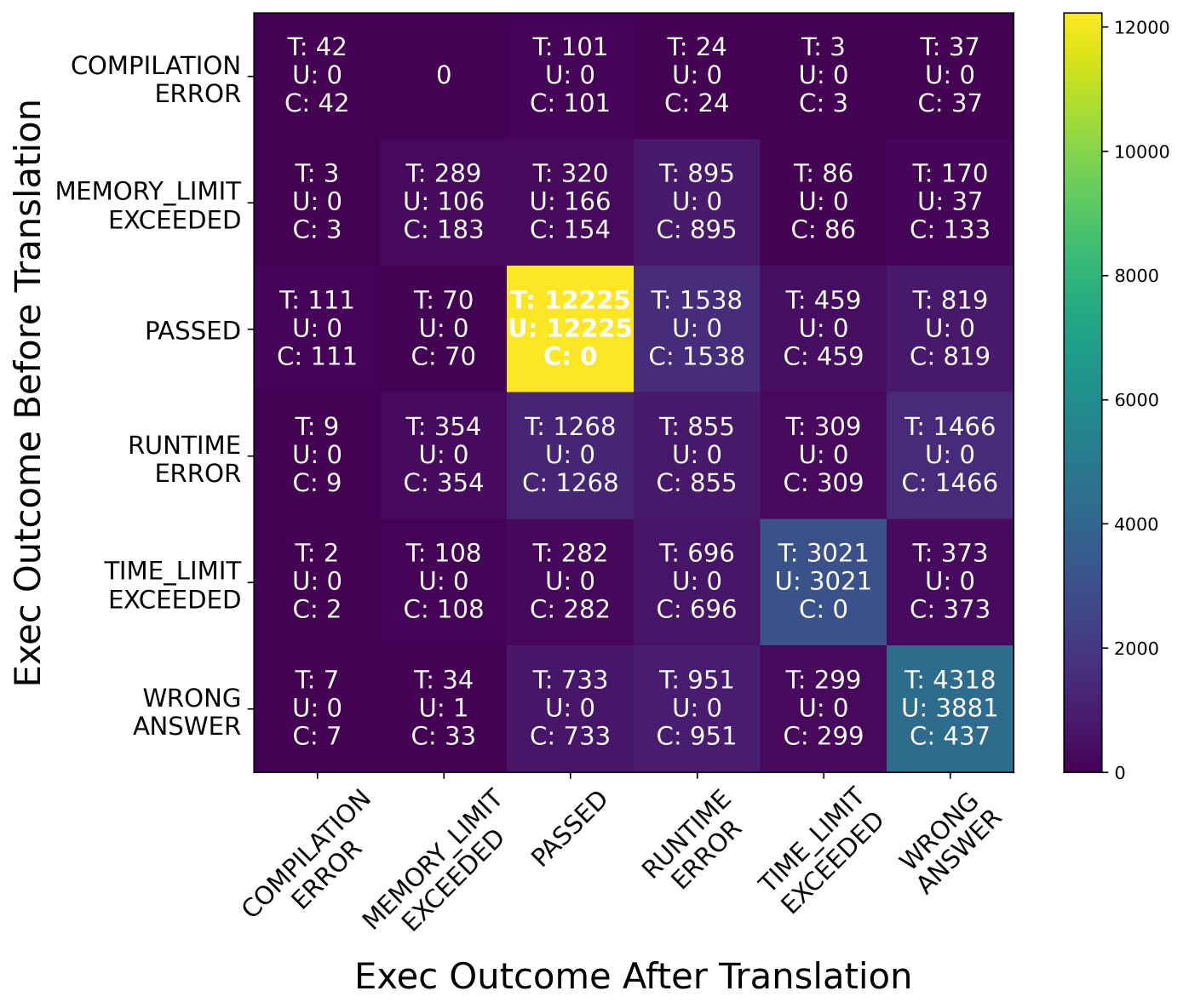}
  \caption{Heatmap of bug execution transition before and after translation (T: total number of test cases, U: test cases where the outcome remains unchanged after translation, C: test cases where the outcome changes after translation).}
  \label{trans}
\vspace{-3mm}
\end{figure}

\subsubsection{Experimental Results for RQ2.2 (Translation Consistency)}
Figure \ref{trans} presents a heatmap of transitions among the six bug categories (as introduced in Section \ref{bm}) before and after translation. For each transition, we compare the test case outcomes of PASSED and WRONG ANSWER to validate the consistency, while the other categories output compiler-specific error messages, resulting mainly in changed outcomes after translation. In particular, two groups of transitions are illustrated in the heatmap, (1) consistent transitions (on the diagonal), and (2) the remaining inconsistent transitions.

\begin{figure}[t]
  \vspace{-3mm}
  \centering\includegraphics[width=.85\linewidth]{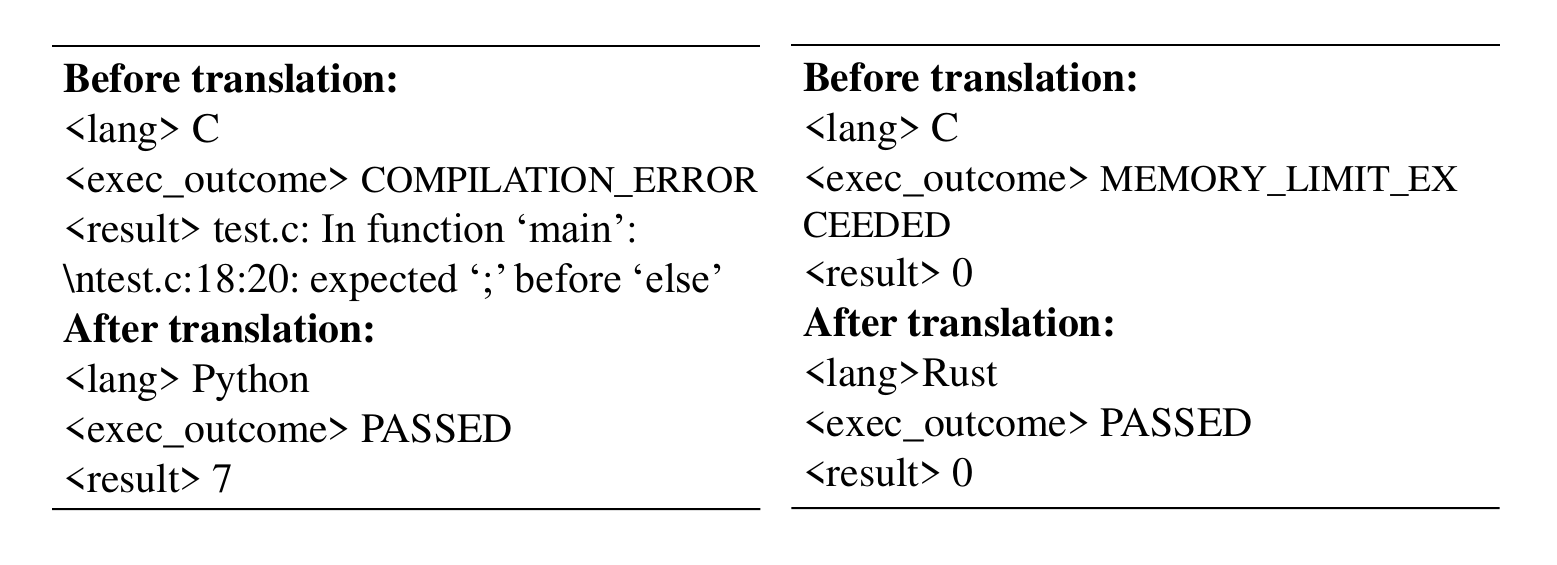}
  \caption{Examples of inconsistent transition.}
  \label{transition-ex}
  \vspace{-3mm}
\end{figure}

For the consistent transitions, we notice that a total of 20750 out of all 32277 cases remain consistent in coarse-grained categories. In particular, the consistent PASSED transitions account for 12225 cases. Moreover, 3881 out of 4318 WRONG ANSWER transitions, and all 3021 TIME LIMIT EXCEEDED transitions remain consistent after translation. Since compilers of different languages produce different error information even for the same category of bugs, the output of COMPILATION ERROR and RUNTIME ERROR is changed after translation, except for MEMORY LIMIT EXCEEDED transitions where C/C++ compilers still return answers for the problem while Java compilers throws an \textit{OutOfMemory} exception, resulting in partial consistency in the outcomes.

\begin{figure}[h]
  \centering
  \begin{subfigure}{\linewidth}
    \centering
    \includegraphics[width=.7\linewidth]{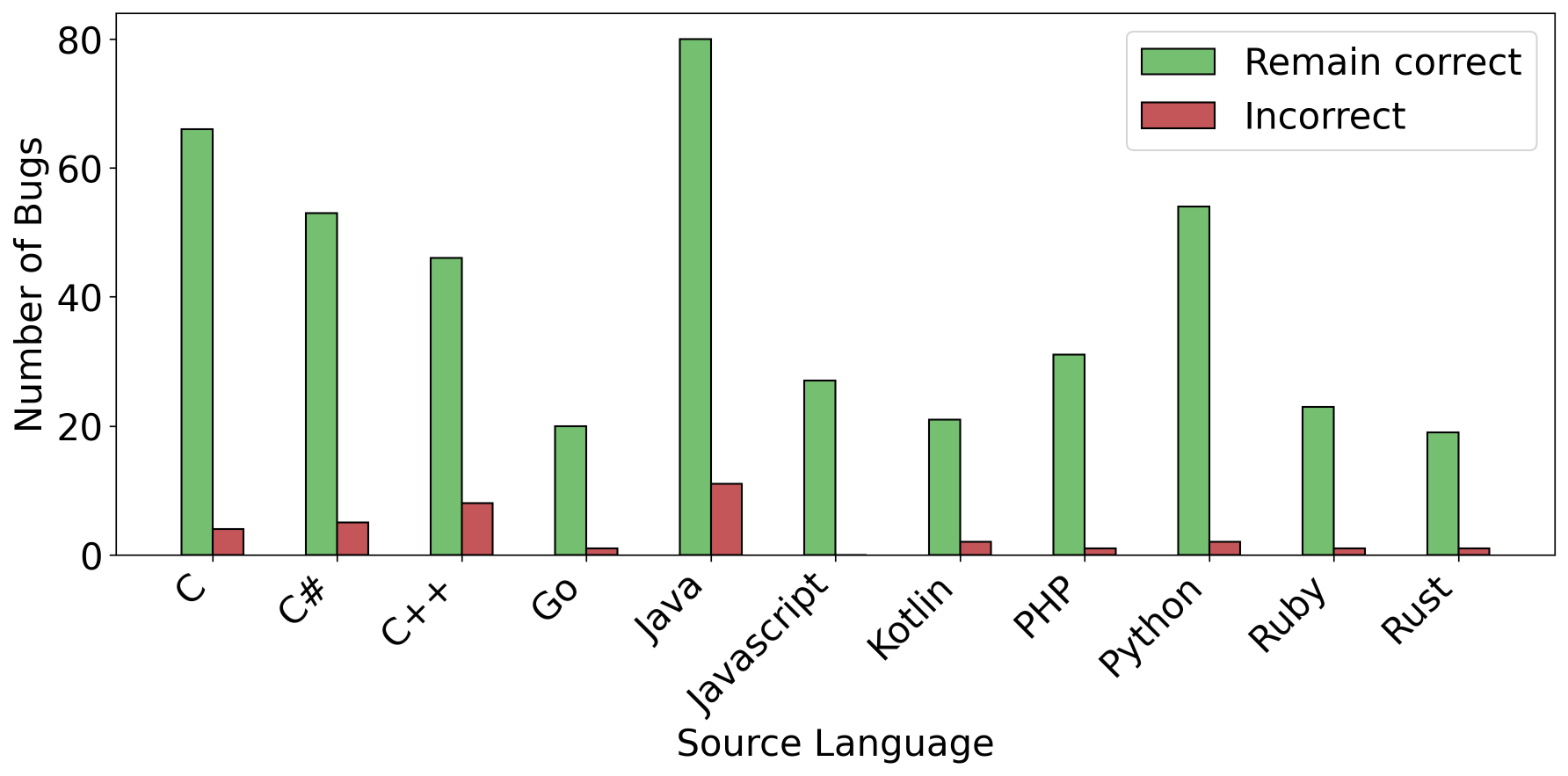}
    \caption{Number of fixed bugs that remain correct or become incorrect after back-translation.}
    \label{bt_bug}
  \end{subfigure}
  \begin{subfigure}{\linewidth}
    \centering
    \includegraphics[width=.7\linewidth]{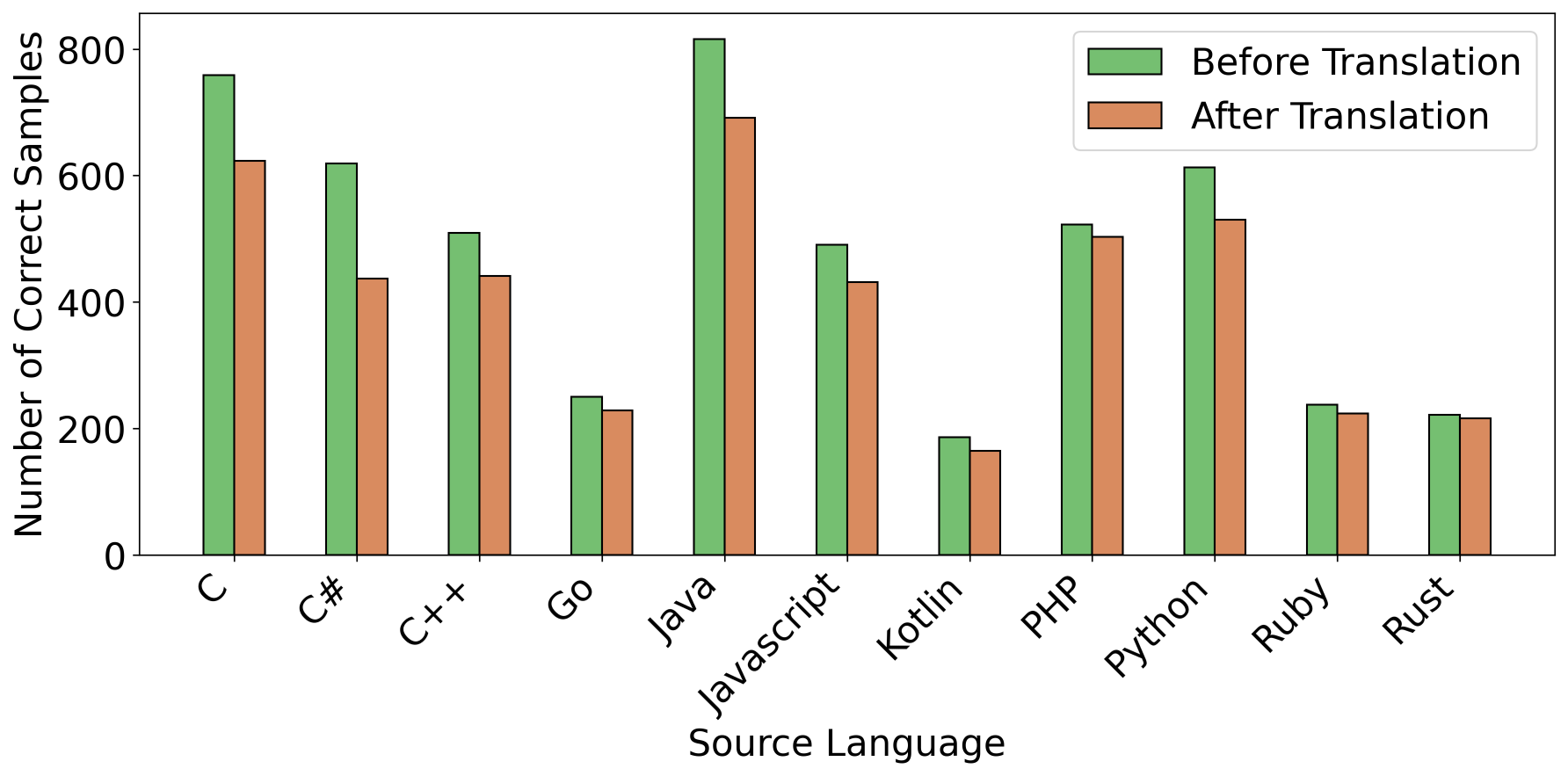}
    \caption{Number of correct samples before and after back-translation.}
    \label{bt_sample}
  \end{subfigure}
  \caption{Validation for back-translation.}
  \label{bt}
\end{figure}
\vspace{-3mm}

First, we notice that most bugs with COMPILATION ERROR can be fixed early in the translation phase, or transformed into other categories of bugs if minor additional issues remain within them. Figure \ref{transition-ex} shows two examples of inconsistent transition. The first example presents a COMPILATION ERROR case of a C bug missing a semicolon "\textit{;}", which is fixed after being translated to Python. Since Python syntax requires no semicolons, this compilation problem is naturally repaired after translation. Such compilation errors can be fixed even when translating to languages that require semicolons, revealing the inherent features of LLM-based code translation. Secondly, language-specific differences such as memory and time usage are also critical contributors to the transition inconsistency. The second example in Figure \ref{transition-ex} illustrates the transition of a MEMORY LIMIT EXCEEDED case of a C bug to a PASSED case in Rust, which consumes less memory resource, thus enabling the repair. A very small subset of cases, where WRONG ANSWER cases are transformed into PASSED ones after translation also contribute to the inconsistency.

Figure \ref{bt_bug} shows the number of fixed bugs that remain correct and those that become incorrect after back-translation. Java experienced the greatest loss with 11 out of 91 fixed bugs incorrectly translated, which remains a small number of bugs compared to its correctly translated bugs. Figure \ref{bt_sample} exhibits the number of correct samples generated for different bugs before and after back-translation, among which C\# suffers from a maximal loss of 182 correct samples, whereas Rust only encounters a minimal loss of six correct samples. Our statistical tests on correct samples before and after back-translation yield a p-value of 0.39 ($>0.05$) and a Cliff's Delta of 0.22, confirming that the loss caused by back-translation is not significant.
\findx{\textbf{[RQ-2] Findings:} (1) The LLM-based reasoning can effectively identify optimal target languages where model exhibits superior repair ability, achieving a minimal average translation path of 2.52. (2) The bugs addressed by \lantern are complex and hard to tackle, with some resolved bugs having median difficulty 400 to 500 larger. (3) Although language-specific errors sometimes introduce inconsistencies, translation generally preserves most bug semantics (with 12225 consistent PASSED cases and 3881 out of 4318 consistent WRONG ANSWER cases) and can even naturally repair certain defects. (4) Back-translation preserves semantic consistency for most repaired code, with an overall loss of only 7.56\% of fixed bugs while 85.95\% of correct samples are preserved.}
\begin{table*}
\fontsize{7.5}{7.5}\selectfont
\centering
\caption{Pass@10 Performance of Different LLMs on xCodeEval}
\label{rq4.2}
\resizebox{.7\linewidth}{!}{
\begin{tblr}{
  width = \linewidth,
  colspec = {Q[165]Q[104]Q[58]Q[58]Q[58]Q[58]Q[58]Q[58]Q[63]Q[58]Q[71]Q[58]Q[58]},
  cells = {c},
  cell{2}{1} = {r=2}{},
  cell{4}{1} = {r=2}{},
  vline{3} = {-}{},
  hline{1-2,4,6} = {-}{},
}
\textbf{LLM}         & \textbf{Approach} & \textbf{C} & \textbf{C\#} & \textbf{C\textbf{++}} & \textbf{Go} & \textbf{Java} & \textbf{JS} & \textbf{Kotlin} & \textbf{PHP} & \textbf{Python} & \textbf{Ruby} & \textbf{Rust} \\
Claude 3.5 Sonnet    & Direct     & 83.28      & 82.68        & 75.30                 & 82.29       & 79.81         & 79.80       & 75.94           & 81.28        & 83.55           & 78.82         & 67.79         \\
                     & LANTERN           & 90.42      & 87.34        & 80.81                 & 90.46       & 85.99         & 86.40       & 92.89           & 99.23        & 90.66           & 85.89         & 86.09         \\
QWen2.5-72B-Instruct & Direct     & 79.06      & 69.50        & 65.90                 & 74.97       & 67.19         & 82.61       & 84.47           & 81.55        & 82.25           & 78.23         & 45.42         \\
                     & LANTERN           & 86.09      & 79.92        & 75.26                 & 85.11       & 76.89         & 88.30       & 88.21           & 93.22        & 88.37           & 86.69         & 72.00         
\end{tblr}
}
\vspace{-3mm}
\end{table*}

\subsection{RQ3: Ablation Study}
\subsubsection{Experimental Design}
We perform an ablation study to investigate the contribution of the key component and design choice including code translation and historical feedback. We aim to determine the impact of code translation on program repair performance and whether incorporating historical feedback assists the LLM in performing accurate reasoning.

\begin{figure}[t]
  \centering
  \includegraphics[width=.8\linewidth]{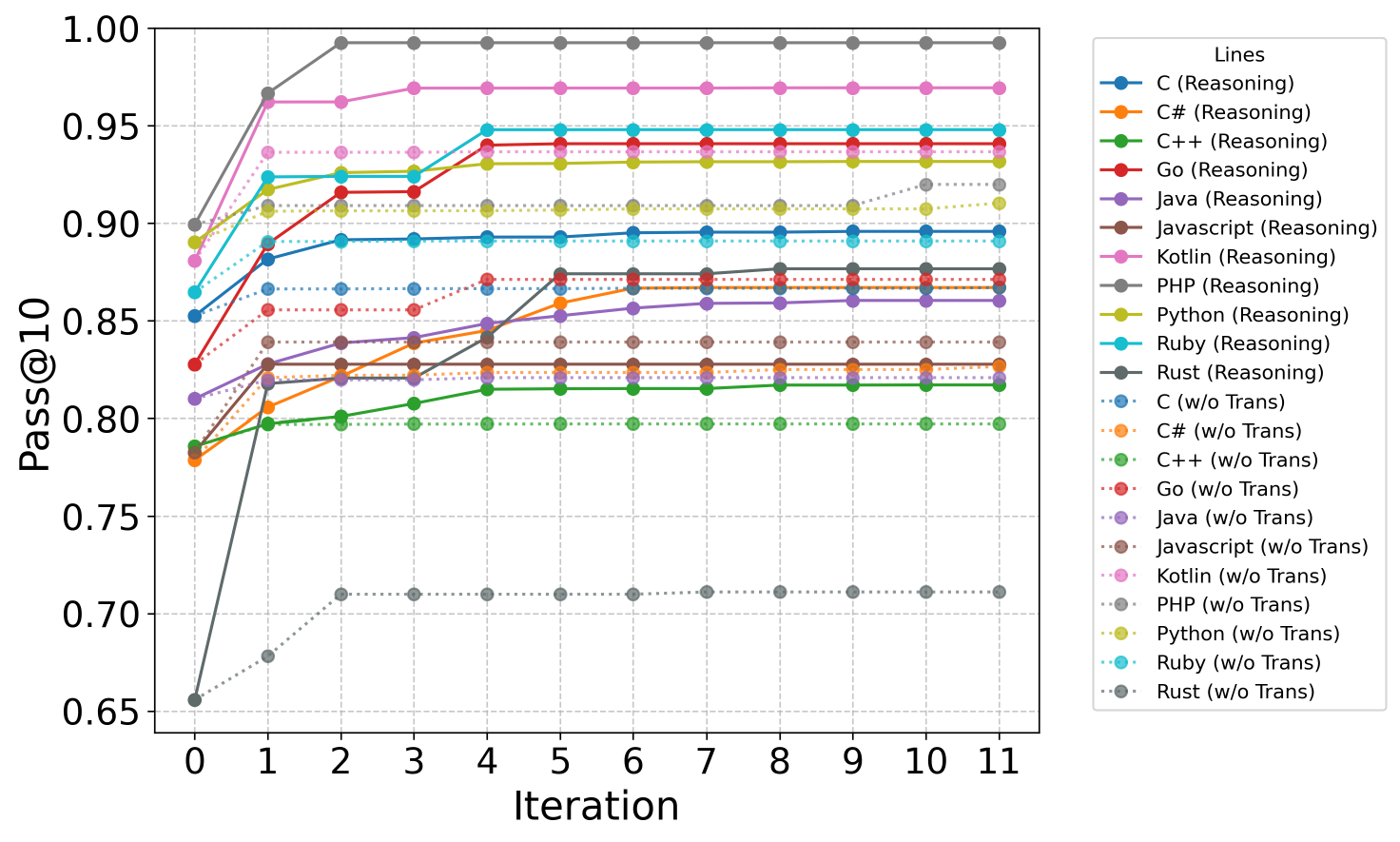}
  \caption{Pass@10 across iterations with and without translation (Reasoning and w/o Trans refer to the reasoning strategy and without translation).}
  \label{notrans}
  \vspace{-0.5cm}
\end{figure}

\begin{figure}[t]
  \centering
  \includegraphics[width=.8\linewidth]{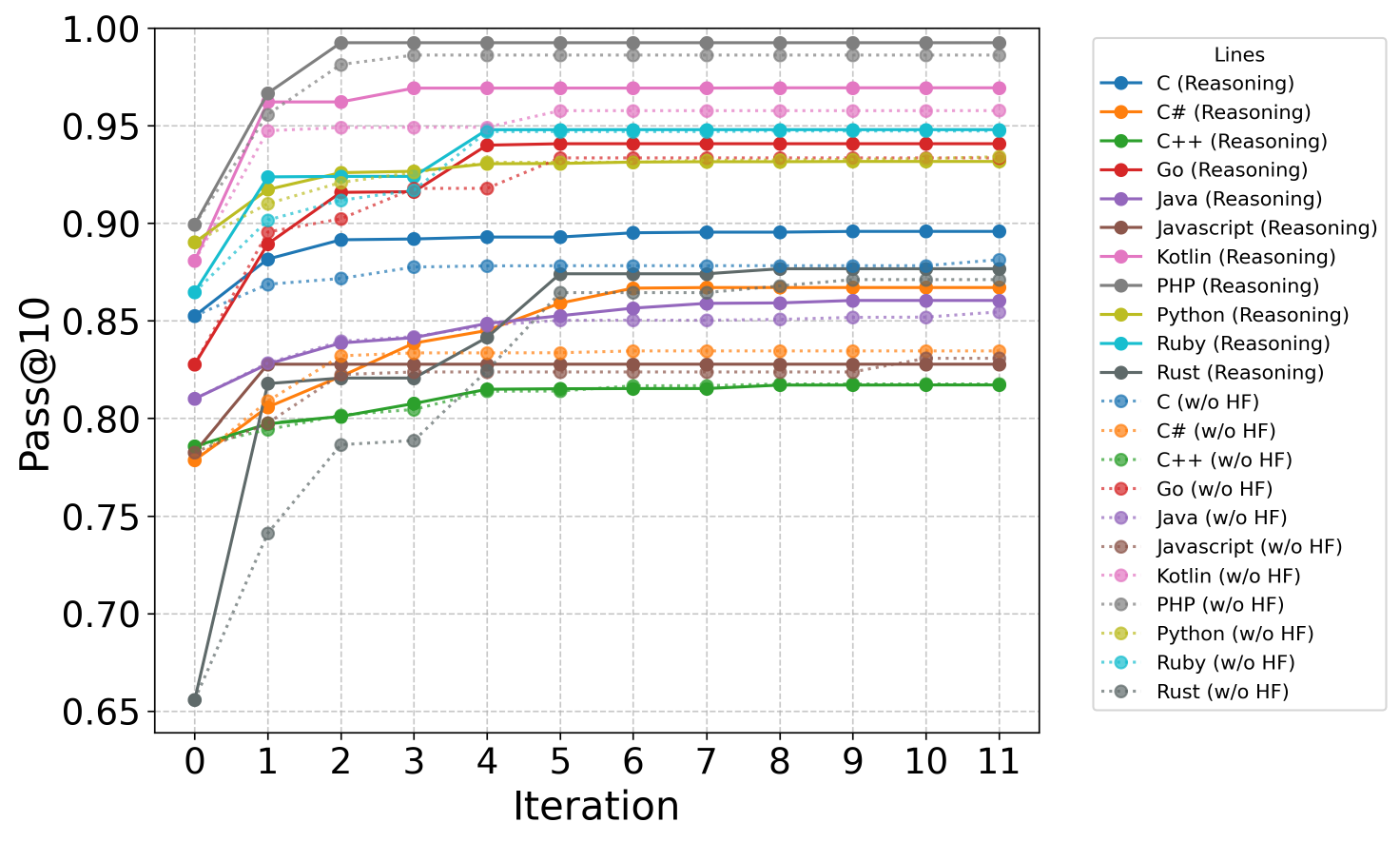}
  \caption{Pass@10 across iterations with/without historical feedback (w/o HF refers to without historical feedback).}
  \label{nohist}
\vspace{-3mm}
\end{figure}

\subsubsection{Experimental Results}
Figure \ref{notrans} exhibits the $Pass@10$ results of the reasoning strategy and direct repair without translation. We notice that repair without translation shows only a slight performance improvement during the initial iteration, after which further fixes become consistently difficult to achieve. Notably, the reasoning strategy achieves a 16.55\% improvement in $Pass@10$ for Rust compared to direct repair, followed by PHP, Go, and Ruby, with improvements of 7.27\%, 6.96\%, and 5.71\%, respectively. Our statistical analysis of the final Pass@10 improvements confirms that translation-based repair significantly outperforms direct repair with a p-value of $1.95E-03$ and Cliff's Delta of 0.42 (medium effect).


Figure \ref{nohist} compares using the reasoning strategy with and without analysis on historical feedback. We see that the final $Pass@10$ performance of the two remains similar, which is expected since both are fundamentally translation-based repair like other strategies. However, the most significant discrepancy manifests during the early iteration cycles, where incorporating historical feedback provides noticeable advantages. For example, in Rust, our approach achieves a $Pass@10$ of 0.82 in the first iteration compared to only 0.74 without historical feedback. Without this feedback, the LLM selects target languages solely based on the bug itself, leading to suboptimal target languages. 

\findx{\textbf{[RQ-3] Findings:} (1) The bug translation component contributes significantly to enhancing program repair effectiveness, with its most substantial impact observed in Rust, achieving a 16.55\% improvement in Pass@10. (2) Identifying appropriate target languages is crucial, as historical feedback facilitates the early convergence of repairs by guiding the selection process toward the most promising options.}

\subsection{RQ4: Generalizability}
\subsubsection{Experimental Design}

In RQ4.1, we evaluate LANTERN on two real-world benchmarks. For Defects4J, we follow ChatRepair's settings, focusing on the most comprehensive single-function subset (255 bugs from v1.2) and the filtered set of 78 bugs from v2.0. While ChatRepair used GPT-3.5, we employ DeepSeek-V3 for a more current comparison. For SWE-bench, we align with AGENTLESS evaluation settings, leveraging their fault localization results and extracting 10-line context windows around bug locations. We use AGENTLESS 1.5 with GPT-4o and employ the same model for fair comparison, measuring performance by resolved instances. Both benchmarks use the same 11 target languages from xCodeEval. 

Specifically, we translate the whole buggy function for Defects4J and the extracted context window (i.e., the bug location with a fixed number of context lines around it) for SWE-bench. In terms of bug repair, we provide the same context (e.g., problem description, buggy code, etc.) following the two baselines, respectively.

For RQ4.2, we investigate LANTERN's generalizability across different models by replicating core xCodeEval experiments using Claude 3.5 Sonnet and Qwen2.5-72B-Instruct, with direct repair as the baseline. This determines whether our translation-based paradigm generalizes across model architectures.

\subsubsection{Experimental Results for RQ4.1 (Real-world Generalizability)}


\begin{table}[t]
\fontsize{7.5}{7.5}\selectfont
\centering
\caption{Performance on Defects4J and SWE-Bench Lite}
\label{rq4.1}
\resizebox{.8\columnwidth}{!}{
\begin{tblr}{
  width = \linewidth,
  colspec = {Q[152]Q[110]Q[110]Q[162]Q[402]},
  cells = {c},
  vline{4} = {-}{},
  hline{1-2,4} = {-}{},
}
\textbf{Approach} & \textbf{D4J1.2} & \textbf{D4J2.0} & \textbf{Approach} & \textbf{SWE-Bench Lite} \\
ChatRepair        & 141              & 54               & AGENTLESS         & 95/300 (31.67\%)                      \\
LANTERN           & \textbf{170}     & \textbf{60}      & LANTERN           & \textbf{110/300 (36.67\%)}            
\end{tblr}
}
\vspace{-3mm}
\end{table}

Table \ref{rq4.1} shows LANTERN's performance on Defects4J and SWE-Bench compared to ChatRepair and AGENTLESS, respectively. On Defects4J 1.2, LANTERN resolves 170 bugs versus ChatRepair's 141 (+29 bugs). For Defects4J 2.0, LANTERN fixes 60 bugs compared to ChatRepair's 54, suggesting that while the source language is only Java, translating the buggy context to a different language (e.g., Python for its expressive syntax or C++ for its low-level control) appears to help the LLM overcome being stuck in a "local optimum" of incorrect repair attempts and produce more effective patches.

On SWE-Bench Lite, LANTERN achieves stronger results, resolving 110/300 instances (36.67\%) versus AGENTLESS's 95/300 (31.67\%). This benchmark requires resolving real-world GitHub issues across multiple files, yet LANTERN succeeds by translating only localized context windows around fault locations. This demonstrates that complete semantic representation of entire projects is unnecessary—focusing on relevant context maintains both scalability and efficiency for large-scale codebases.

The results indicate LANTERN's effectiveness extends beyond traditionally under-represented languages. Even for widely-used languages like Java and Python, LANTERN transforms difficult instances into more tractable problems by leveraging different programming languages' unique strengths, suggesting it functions as a general-purpose strategy for enhancing problem-solving capabilities on complex bugs.

\subsubsection{Experimental Results for RQ4.2 (Model Generalizability)}
Table \ref{rq4.2} presents Pass@10 results for Claude 3.5 Sonnet and QWen2.5-72B-Instruct on xCodeEval. LANTERN elevates Claude's performance across every language, with massive boosts in Kotlin (+16.94\%), PHP (+17.95\%), and Rust (+18.30\%), demonstrating that even frontier proprietary models benefit significantly from cross-language translation.

Results with QWen2.5-72B-Instruct further highlight LANTERN's power for open-source models. QWen's baseline shows significant variance, particularly struggling with languages requiring strict compilation or complex type systems like Rust (45.42\%) and C\# (69.50\%). LANTERN provides dramatic improvements on Rust (+26.58\%), C\# (+10.42\%), Go (+10.13\%), and PHP (+11.67\%), revealing that LANTERN can elevate strong open-source models to achieve performance competitive with or superior to highly-capable proprietary models. For instance, QWen with LANTERN achieves 72.00\% on Rust, surpassing Claude's direct repair score of 67.79\%.

\findx{\textbf{[RQ-4] Findings:} (1) LANTERN outperforms ChatRepair and AGENTLESS on the challenging real-world benchmarks, which demonstrates strong generalizability to real-world issue resolution, revealing its cross-language translation paradigm is effective for complex, real-world software defects, including those that require understanding context across multiple files. (2) LANTERN demonstrates model-agnostic generalizability across different LLMs. It consistently and significantly enhances the repair capabilities of both state-of-the-art closed-source and leading open-source LLMs.}

\vspace{-3mm}
\section{Threats to Validity}
\textbf{Internal.} 
The internal threat arises from the historical feedback, where as iterations progress and more historical feedback is generated, subsequent reasoning may increasingly bias toward previously successful languages. We mitigate this by limiting that each language can only be selected once, and comparing the reasoning strategy against both greedy and random strategies, showing that the feedback-based strategy outperforms the others. 

\noindent\textbf{Construct.} We primarily use Pass@10 for evaluation, which might not capture all repair aspects. To address this, we incorporate multiple additional ranking metrics to provide a more comprehensive assessment. Another threat comes from potential semantic inconsistencies introduced during translation. We mitigate this by validating the semantic consistency for both bug translation and fixed code back-translation through extensive test cases.

\noindent\textbf{External.} We evaluate on a large number of bugs (xCodeEval), demonstrating effectiveness across 11 diverse programming languages and bug types. Furthermore, our significant improvements on less common languages like Rust suggest that the approach generalizes beyond mainstream languages. Our approach can also potentially be applied to repository-level codebases. The pipeline's design can also be extended to additional programming languages and models. 

\vspace{-3mm}

\section{Conclusion}
In this paper, we presented a novel program repair approach, LAN-TERN, which leverages cross-language code translation with multi-agent iterative refinement to fix bugs by translating buggy code to languages where the LLM demonstrates stronger repair capabilities based on the bug characteristics and historical feedback. Evaluation on xCodeEval with 5,068 bugs across 11 programming languages shows that our approach can significantly enhance the repair capability of the LLM, with notable improvements in less common languages like Rust (with a 22.09\% increase in $Pass@10$). Our results demonstrate the potential of cross-language program repair in effectively extending the repair capabilities of the LLM, revealing new opportunities for agent-guided automated program repair.

\bibliographystyle{ACM-Reference-Format}
\bibliography{main}

\end{document}